\documentclass[prb,aps,twocolumn,superscriptaddress]{revtex4}
\usepackage{graphicx}
\usepackage{amsmath}
\usepackage{amssymb}
\usepackage{dsfont}
\usepackage{bm}
\usepackage[hidelinks]{hyperref}
\usepackage{multirow}
\usepackage{appendix}
\usepackage{color}
\usepackage{placeins}

\newcommand{\braket}[2]{\left\langle #1 | #2 \right\rangle}
\newcommand{\bra}[1]{\left\langle#1\right|}
\newcommand{\ket}[1]{\left|#1\right\rangle}

\begin{document}
\title{Higher-Order Topology, Monopole Nodal Lines, and the Origin of Large Fermi Arcs in Transition Metal Dichalcogenides XTe$_2$ (X$=$Mo,W)}
\author{Zhijun Wang}
\thanks{These authors contributed equally to this work.}
\affiliation{Beijing National Laboratory for Condensed Matter Physics,
and Institute of Physics, Chinese Academy of Sciences, Beijing 100190, China}
\affiliation{University of Chinese Academy of Sciences, Beijing 100049, China}
\affiliation{Department of Physics,
Princeton University,
Princeton, NJ 08544, USA}

\author{Benjamin J. Wieder}
\thanks{These authors contributed equally to this work.}
\affiliation{Department of Physics,
Princeton University,
Princeton, NJ 08544, USA}

\author{Jian Li}
\affiliation{School of Science,
Westlake University, 18 Shilongshan Road,
Hangzhou 310024, China
            }

\author{Binghai Yan}
\affiliation{Department of Condensed Matter Physics, 
Weizmann Institute of Science, 
Rehovot, 7610001, Israel
            }

\author{B. Andrei Bernevig}
\affiliation{Department of Physics,
Princeton University,
Princeton, NJ 08544, USA}
\affiliation{Dahlem Center for Complex Quantum Systems and Fachbereich Physik,
Freie Universit{\"a}t Berlin, Arnimallee 14, 14195 Berlin, Germany
	}
\affiliation{Max Planck Institute of Microstructure Physics, 
06120 Halle, Germany
}

\date{\today}
\begin{abstract}
In recent years, transition metal dichalcogenides (TMDs) have garnered great interest as topological materials.  In particular, monolayers of centrosymmetric $\beta$-phase TMDs have been identified as 2D topological insulators (TIs), and bulk crystals of noncentrosymmetric $\gamma$-phase MoTe$_2$ and WTe$_2$ have been identified as type-II Weyl semimetals.  However, ARPES and STM probes of these semimetals have revealed huge, ``arc-like'' surface states that overwhelm, and are sometimes mistaken for, the much smaller topological surface Fermi arcs of bulk type-II Weyl points.  In this letter, we calculate the bulk and surface electronic structure of both $\beta$- and $\gamma$-MoTe$_2$.  We find that $\beta$-MoTe$_2$ is in fact a $\mathbb{Z}_{4}$-nontrivial higher-order TI (HOTI) driven by double band inversion and exhibits the \emph{same} surface features as $\gamma$-MoTe$_2$ and $\gamma$-WTe$_2$.  We discover that these surface states are not topologically trivial, as previously characterized by the research that differentiated them from the Weyl Fermi arcs, but rather are the characteristic split and gapped fourfold Dirac surface states of a HOTI.  In $\beta$-MoTe$_2$, this indicates that it would exhibit helical pairs of hinge states if it were bulk-insulating, and in $\gamma$-MoTe$_2$ and $\gamma$-WTe$_2$, these surface states represent vestiges of HOTI phases without inversion symmetry that are nearby in parameter space.  Using nested Wilson loops and first-principles calculations, we explicitly demonstrate that when the Weyl points in $\gamma$-MoTe$_2$ are annihilated, which may be accomplished by symmetry-preserving strain or lattice distortion, $\gamma$-MoTe$_2$ becomes a \emph{non-symmetry-indicated, noncentrosymmetric} HOTI.  We also show that, when the effects of SOC are neglected, $\beta$-MoTe$_2$ is a nodal-line semimetal with $\mathbb{Z}_{2}$-nontrivial monopole nodal lines (MNLSM).  This finding confirms that MNLSMs driven by double band inversion are the weak-SOC limit of HOTIs, implying that MNLSMs are higher-order topological \emph{semimetals} with flat-band-like hinge states, which we find to originate from the corner modes of  2D ``fragile'' TIs. 
\end{abstract}

\maketitle

Over the past decade, the number of topological insulating (TI) and semimetallic (SM) phases identified in real materials has grown immensely.  With the discovery of increasingly intricate TIs and SMs, such as higher-order TIs (HOTIs)~\cite{multipole,WladTheory,HOTIBernevig,HOTIChen,HigherOrderTIPiet,HigherOrderTIPiet2,DiracInsulator,FanHOTI,EzawaMagneticHOTI,ZeroBerry,WiederAxion,KoreanAxion} and unconventional fermion SMs~\cite{DDP,NewFermions,KramersWeyl,RhSiArc,CoSiArc,SteveMagnet}, previously overlooked, but readily accessible compounds, including bismuth~\cite{HOTIBismuth} and chiral B20~\cite{AlPtObserve,CoSiObserve1,CoSiObserve2,CoSiObserve3,PdGaObserve} crystals, have been experimentally verified as topologically nontrivial.  In this letter, we extend the theory and experimental applicability of higher-order topology by recognizing that the XTe$_2$ (X$=$Mo,W) family~\cite{XTe2Structures,CarlSynthesis1,CarlSynthesis2,OtherSynthesis1,OtherSynthesis2,LeslieBobTitanic} of transition metal dichalcogenides (TMDs), a large, well-studied, and readily synthesizable class of materials, are HOTIs, and not topologically trivial.  XTe$_2$ TMDs, which exhibit a trivial magnetoelectric polarizability~\cite{WilczekAxion,VDBAxion,VDBHOTI,QHZ,AndreiInversion,AshvinAxion1,AshvinAxion2,WuAxionExp,WiederAxion}, therefore provide an intriguing experimental platform for examining topological response effects beyond the magnetoelectric effect~\cite{TaylorTCIResponse,WiederDefect,ZeroHall,BarryConvo,BarryPrep}.  We also draw connections between an exotic class of nodal-line SMs (NLSMs)~\cite{YoungkukLineNode,FangWithWithout,YoungkukMonopole} and the recently introduced notion of ``fragile'' topology~\cite{AshvinFragile,JenFragile1,JenFragile2,HingeSM,AdrianFragile,WiederAxion}, leading to the discovery of fractionally charged corner modes in fragile TIs.  Our findings further establish the seemingly esoteric concept of higher-order topology as \emph{crucial} for characterizing topological transport and response effects in everyday materials.

All of the spinful TIs discovered to date represent the gapped, spin-orbit coupled (strong-SOC) limits of gapless topological (SM) phases without SOC.  This intrinsic link between gapped and gapless phases has defined topological condensed matter physics since the recognition that graphene~\cite{GrapheneDirac,semenoff,meleDirac} and HgTe gap into $\mathbb{Z}_{2}$ TIs~\cite{KaneMeleZ2,AndreiTI} under the introduction of SOC~\cite{CharlieTI}.  As the number of known topological SMs has increased~\cite{WeylReview,BinghaiReview,NaDirac,SchnyderDirac,NagaosaDirac,ZJDirac,SyDiracSurface,borisenkoDirac,SteveDirac,AshvinWeyl1,soluyanov_type-ii_2015,wang_mote2:_2016,sun_prediction_2015,DDP,NewFermions,SteveMagnet,KramersWeyl,RhSiArc,CoSiArc}, the number of known topological (crystalline) insulators realized by gapping them has kept pace~\cite{LiangTCI,HsiehTCI,SCZAxion,Steve2D,HourglassInsulator,DiracInsulator,ChenRotation}.  In one particularly simple example, an SM phase with a ring of linearly-dispersing degeneracies, known as a ``nodal-line'' (NL), can occur in a weak-SOC crystal with only inversion ($\mathcal{I}$) and time-reversal ($\mathcal{T}$) symmetries~\cite{YoungkukLineNode,XiLineNode,SchnyderDrumhead}.  NLs may be created and annihilated at single, time-reversal-invariant (TRIM) points in the Brillouin zone (BZ) by inverting bands with opposite parity ($\mathcal{I}$) eigenvalues~\cite{YoungkukLineNode}, such that the number of NLs is constrained by the \emph{same} Fu-Kane parity criterion~\cite{FuKaneMele,FuKaneInversion} that indicates 3D TI phases in strong-SOC crystals.  This recognition has driven the rapid identification of candidate NLSMs, including Ca$_3$P$_2$~\cite{LeslieLineCa}, Cu$_3$(Pd,Zn)N~\cite{YoungkukLineNode,XiLineNode}, and 3D graphene networks~\cite{GrapheneNetworkLine}, all of which exhibit characteristic nearly-flat-band ``drumhead'' surface states.  Crucially, it also implies that weak-SOC NLSMs gap directly into 3D TIs upon the introduction of $\mathcal{I}$-symmetric SOC~\cite{YoungkukLineNode}.

\begin{figure}
\includegraphics[width=0.65\linewidth]{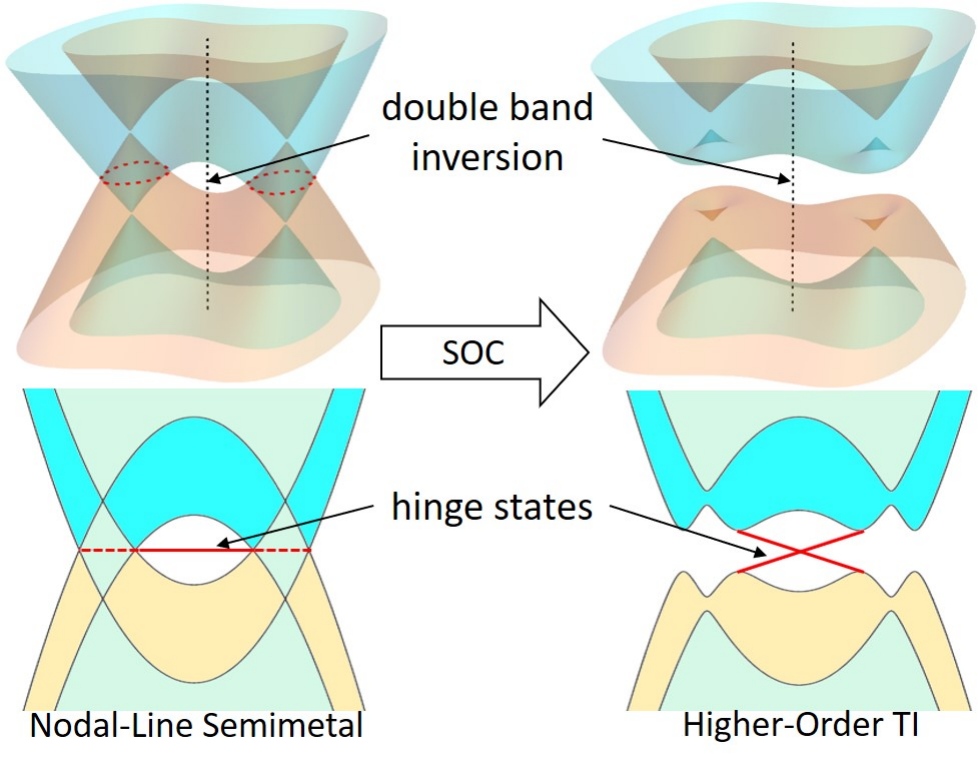}
\caption{When two pairs of degenerate bands with positive parity eigenvalues and two pairs with negative parity eigenvalues are inverted at a TRIM point~\cite{HOTIBernevig,HOTIBismuth}, the occupied bands cannot be expressed as a linear combination of EBRs~\cite{ZakBandrep1,ZakBandrep2,QuantumChemistry,Bandrep1,Bandrep2,Bandrep3,JenFragile1} and the $\mathbb{Z}_{4}$ topological index~\cite{ChenTCI,AshvinIndicators,AshvinTCI,EslamInversion} is changed by 2.  In a $\mathcal{T}$-symmetric crystal with vanishing SOC, this process may nucleate a pair of Dirac nodal lines with nontrivial monopole charge (MNLs)~\cite{YoungkukMonopole} (dashed lines in left panel).  On the 1D hinges, the projections of the MNLs are spanned by nearly-flat bands (an explicit model is provided in SA~\ref{sec:TBmodel}).  These hinge states represent an example of higher-order topology in a bulk-gapless system: they are the $d-2$-dimensional generalization of drumhead surface states, and are the spinless analogs of the hinge states recently predicted in spinful Dirac SMs~\cite{HingeSM,TaylorToy}.  When $\mathcal{I}$-symmetric SOC is introduced, the MNLSM will necessarily gap into a HOTI if all other bands are uninverted, and the flat-band hinge states will open into helical pairs spanning the bulk and surface gaps.  HOTIs driven by this ``double band inversion'' (DBI) include bismuth~\cite{HOTIBismuth} and, as shown in this letter, $\beta$-MoTe$_{2}$ (Fig.~\ref{fig:Fig2}(d)).}
\label{fig:doubleBand}
\end{figure}

Very recently, fundamentally distinct $\mathcal{I}$- and $\mathcal{T}$-symmetric SMs and insulators have been proposed that escape this paradigm.  In~\onlinecite{FangWithWithout}, Fang \emph{et al.} introduced a second kind of weak-SOC NL, which, unlike the previous example, can \emph{only} be removed by pairwise annihilation.  Though the mechanisms underpinning the protection and identification of these ``monopole-charged'' NLs (MNLs) have been explored in detail~\cite{YoungkukMonopole,AdrianMonopole,SigristMonopole}, MNLs have thus far only been proposed in magnonic systems~\cite{ChenMagnon} and 3D graphdiyne~\cite{GraphdiyneExp,YoungkukMonopole}.  Recent works have also identified 3D ``higher-order'' TIs~\cite{multipole,WladTheory,HOTIBernevig,HOTIChen,HigherOrderTIPiet,HigherOrderTIPiet2,DiracInsulator,FanHOTI,EzawaMagneticHOTI,ZeroBerry,WiederAxion,KoreanAxion} stabilized by only $\mathcal{I}$ and $\mathcal{T}$ symmetries~\cite{HOTIBernevig,HOTIBismuth,ChenTCI,AshvinIndicators,AshvinTCI,EslamInversion}.  Notably, 3D HOTIs exhibit gapped 2D surfaces and gapless 1D hinges with characteristic helical modes~\cite{WladTheory,HOTIBernevig,HOTIChen}, which represent the domain wall states between 2D faces with oppositely gapped fourfold Dirac fermions~\cite{DiracInsulator,AshvinTCI}.  By enumerating the parity eigenvalues of trivial (atomic) insulators, whose occupied bands define ``elementary'' band representations (EBRs)~\cite{ZakBandrep1,ZakBandrep2,QuantumChemistry,Bandrep1,Bandrep2,Bandrep3,JenFragile1}, it can be shown that the $\mathbb{Z}_{2}$ Fu-Kane criterion should be promoted to a $\mathbb{Z}_{4}$ index that captures both TIs and HOTIs~\cite{HOTIBernevig,HOTIBismuth,ChenTCI,AshvinIndicators,AshvinTCI}.  Using EBRs~\cite{QuantumChemistry}, HOTI phases have been identified in systems with \emph{double band inversion} (DBI) (Fig.~\ref{fig:doubleBand}), most notably in rhombohedral bismuth crystals~\cite{HOTIBismuth}.

\begin{figure}
\includegraphics[width=1.0\linewidth]{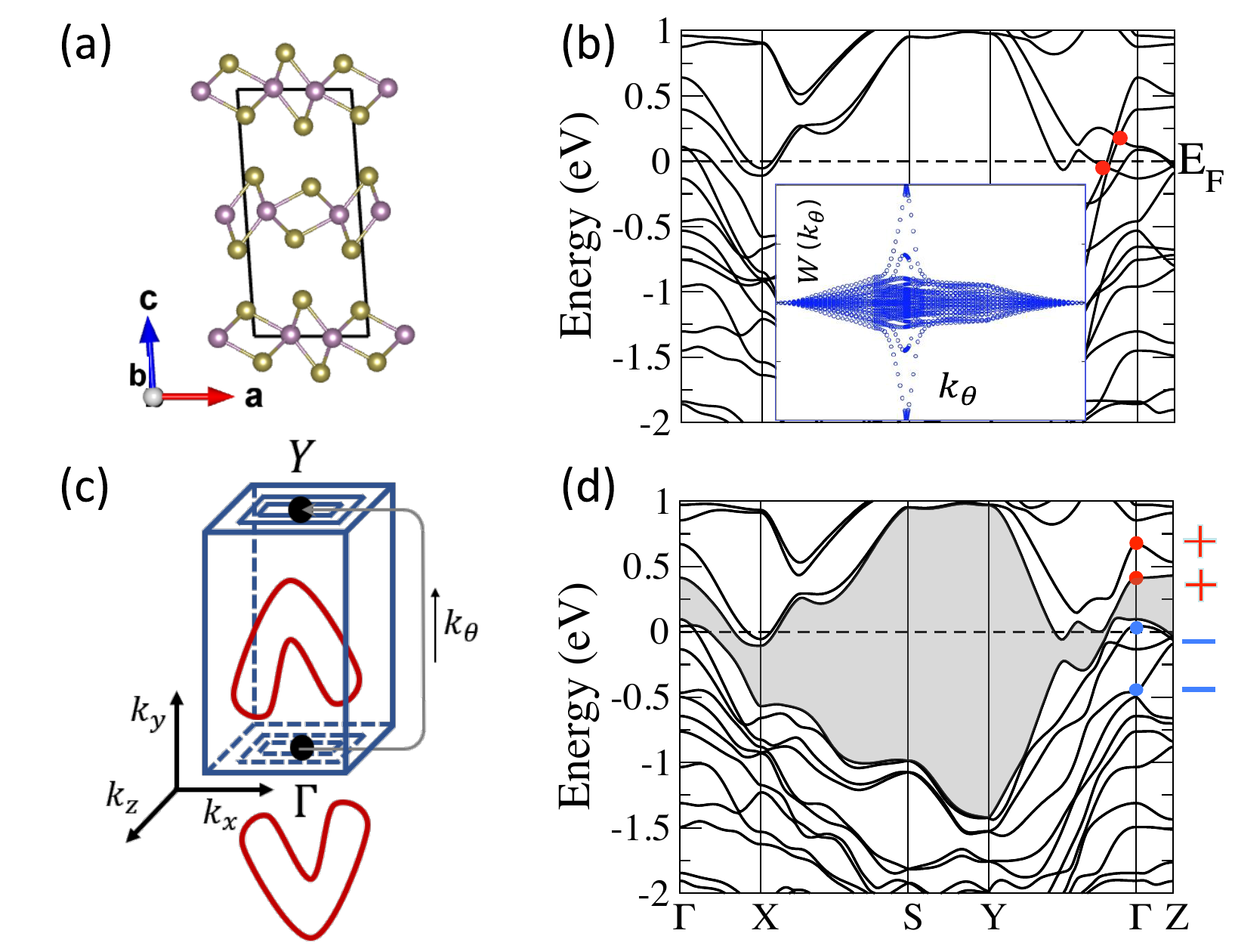}
\caption{(a) The monoclinic lattice of $\beta$- (1T'-) MoTe$_2$~\cite{XTe2Structures} in SG 11 $P2_{1}/m$.  (b,d) Bulk bands of $\beta$-MoTe$_2$ calculated without and with the effects of SOC incorporated, respectively (details provided in SA~\ref{sec:DFT}).  DBI occurs about the $\Gamma$ point, as indicated by the parity eigenvalues in (d).  (c) When SOC is neglected, a $\mathcal{T}$-reversed pair of irregularly shaped NLs forms at $E_{F}$, intersecting $Y\Gamma$ (red dots in (b)) between $k_{y}= \pm 0.06\ (2\pi/b)$ and $k_{y}= \pm 0.19\ (2\pi /b)$, where $b=4.369$ \AA\ is the lattice spacing along the $\vec{b}$ lattice vector~\cite{Brown1966}.  We surround one of the NLs with a closed, tetragonal prism and calculate the Wilson loop around $k_{y}$-normal squares as a function of the polar momentum $k_{\theta}$ (exact coordinates provided in SA~\ref{sec:DFTmonopole}); we observe $\mathbb{Z}_{2}$-nontrivial winding (inset panel in (b)), indicating a nontrivial monopole charge~\cite{YoungkukMonopole,AdrianMonopole,SigristMonopole}.  (d) When SOC is introduced, a gap near $E_{F}$ develops at all crystal momenta (gray shaded region) with the $\mathbb{Z}_{4}$ parity index (Table~\ref{tb:z4}) of a HOTI~\cite{ChenTCI,AshvinIndicators,AshvinTCI,EslamInversion}.}
\label{fig:Fig2}
\end{figure}

Employing this $\mathbb{Z}_{4}$ index and Wilson loops~\cite{Fidkowski2011,AndreiXiZ2,ArisInversion,ArisBerry,Cohomological}, we identify the TMD~\cite{XTe2Structures} $\beta$- (1T'-) MoTe$_2$ (space group (SG) 11 $P2_{1}/m$) as a HOTI with large, gapped, arc-like surface states, and explicitly show that in the absence of SOC, it forms an NLSM with MNLs.  This cements the suggestion, introduced in~\onlinecite{ZhidaSemimetals}, that MNLSMs formed from DBI are the weak-SOC limit of HOTIs, in analogy to the earlier recognition~\cite{YoungkukLineNode} that monopole-trivial NLSMs are the weak-SOC limit of 3D TIs.  It has been shown that the noncentrosymmetric $\gamma$- (Td-) phases of XTe$_2$ (X$=$Mo,W) (SG 31 $Pmn2_{1}$), previously identified as type-II (tilted) Weyl (semi)metals~\cite{wang_mote2:_2016,soluyanov_type-ii_2015,sun_prediction_2015}, exhibit the same large surface states as $\beta$-MoTe$_2$.  These states were previously identified as topologically trivial~\cite{sun_prediction_2015,TamaiMoTe2,WTe2Arpes2,WTe2Arpes3,WTe2Arpes4,MoTe2Arpes1,MoTe2Arpes2,MoTe2Arpes4,WTe2STM,WTe2QPI2018,MoTe2STM,WTe2Arpes1,MoTe2Arpes3} (Fig.~\ref{fig:surf}(d)), as the actual topological Fermi arcs from the Weyl points are considerably shorter~\cite{soluyanov_type-ii_2015,sun_prediction_2015,TamaiMoTe2,MoTe2Arpes1,WTe2Arpes2,WTe2Arpes4}.  However, using nested Wilson loops and first-principles calculations (SA~\ref{sec:DFTgappedGamma}), we explicitly demonstrate that $\gamma$-MoTe$_2$, which exhibits the same band ordering and surface states as $\beta$-MoTe$_2$, transitions into a \emph{non-symmetry-indicated} HOTI when its narrowly separated Weyl points are annihilated.  Therefore, the large surface states in $\gamma$-XTe$_2$ are not trivial, but rather are vestiges of a nearby HOTI phase, and originate from DBI, like those in $\beta$-MoTe$_2$.

TMDs are a class of readily synthesizable~\cite{XTe2Structures,CarlSynthesis1,CarlSynthesis2,OtherSynthesis1,OtherSynthesis2,LeslieBobTitanic} layered materials.  Originally highlighted for the semiconducting bandgap of exfoliated monolayers~\cite{MoS2Transistor}, TMDs have recently been recognized as topological materials -- quasi-2D samples of $\beta$-phase TMDs have been identified as 2D TIs~\cite{LiangTMD,TMDTI,WTe2TI}, and 3D samples of $\gamma$-XTe$_2$ have been identified in theory~\cite{wang_mote2:_2016,soluyanov_type-ii_2015,sun_prediction_2015} and experiment~\cite{BulkMoTe2,BulkWTe21,BulkWTe22} as type-II Weyl SMs.  We first focus on MoTe$_2$, and then generalize our findings to the isostructural phases of WTe$_2$.

\begin{figure}
\includegraphics[width=0.9\linewidth]{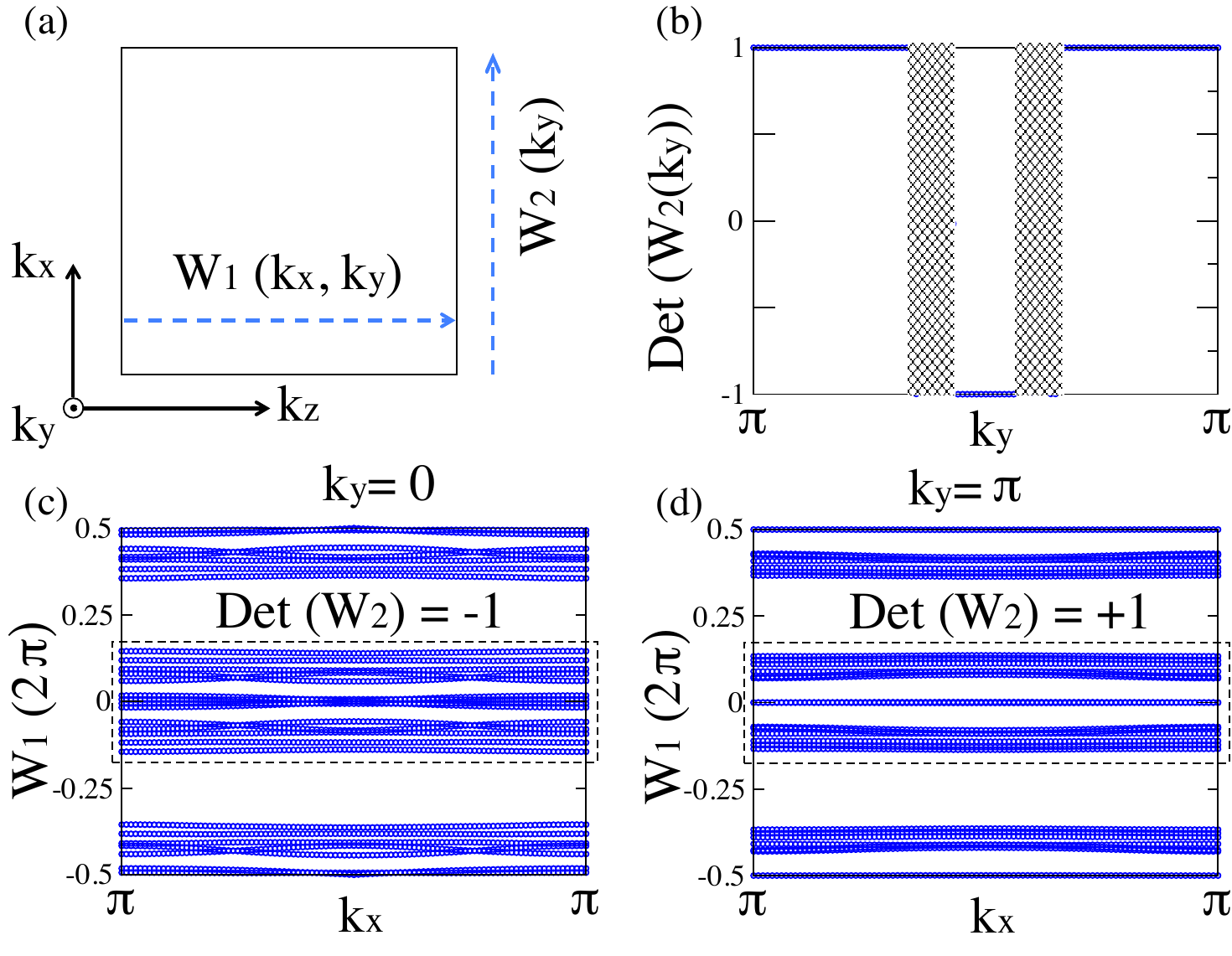}
\caption{(a) Bulk Wilson loops calculated for $\beta$-MoTe$_2$ from first principles in the absence of SOC (SA~\ref{sec:DFTnested}).  At values of $k_{y}$ away from the MNLs in Fig.~\ref{fig:Fig2}, there is a large gap in the $z$-directed Wilson loop spectrum $W_{1}(k_{x},k_{y})$ between $\theta_{1}/2\pi \approx \pm 0.25$; representative examples are shown in (c,d) for $k_{y} = 0,\pi$, respectively.  (b) The determinant of the \emph{nested} Wilson matrix $W_{2}(k_{y})$ calculated over the Wilson bands between $\theta_{1}=\pm \pi/2$~\cite{multipole,WladTheory,HOTIBernevig,HingeSM} is quantized at $\pm 1$ by the antiunitary symmetry $(\mathcal{I}\times\tilde{\mathcal{T}})^{2}=+1$ (SA~\ref{sec:WilsonDefs}), and jumps as it passes over an MNL, indicating that $k_{y}$-indexed planes above and below the MNL are topologically distinct~\cite{multipole,WladTheory,HOTIBernevig}.}
\label{fig:wloop}
\end{figure}

MoTe$_2$ can crystallize in two distinct structures at room temperature: the hexagonal $\alpha$ (2H) phase (SG 194 $P6_{3}/mmc$) and the distorted monoclinic $\beta$ phase~\cite{XTe2Structures,CarlSynthesis1,CarlSynthesis2,MoTe2STM} (Fig.~\ref{fig:Fig2}(a)).  When $\beta$-MoTe$_2$ is further cooled below 250 K, it transitions into the noncentrosymmetric $\gamma$ phase~\cite{XTe2Structures,PhaseTransitionNPJ,MoTe2STM}.  Using first-principles calculations detailed in SA~\ref{sec:DFTmethod}, we calculate the electronic structure of $\beta$-MoTe$_2$ with and without the effects of SOC incorporated (Fig~\ref{fig:Fig2}(d,b), respectively).  $\beta$-MoTe$_2$ exhibits DBI (Fig.~\ref{fig:doubleBand}) at $\Gamma$ as a consequence of the $\beta$-phase lattice distortion (Fig.~\ref{fig:Fig2}(a)).  When SOC is neglected, a $\mathcal{T}$-reversed pair of topological NLs~\cite{YoungkukLineNode,FangWithWithout} forms, intersecting $Y\Gamma$ in an irregular, 3D shape with a significant pucker in the $k_{y}$ direction (schematically depicted in Fig.~\ref{fig:Fig2}(c)).  The NLs (red dots in Fig.~\ref{fig:Fig2}(b)) represent the only crossing points between the bands at $E_{F}$ (taking the direct gap to lie above $N=28$ spin-degenerate pairs of bands).  As prescribed in~\onlinecite{YoungkukMonopole,AdrianMonopole,SigristMonopole}, we surround each NL with a closed surface and calculate the Wilson loop (holonomy) matrix~\cite{Fidkowski2011,AndreiXiZ2,ArisInversion,ArisBerry,Cohomological} over the lower $N$ bands as a function of the polar momentum $k_{\theta}$ (Fig.~\ref{fig:Fig2}(c) and SA~\ref{sec:DFTmonopole}).  This Wilson spectrum exhibits the characteristic winding of an MNL~\cite{YoungkukMonopole,AdrianMonopole,SigristMonopole}.

\begin{table}[t]
\begin{center}
\begin{tabular}{c|cccccccc}
TRIM & $\Gamma$ & X & Y & Z & S & T & U & R \\
\hline
$n_{-}(\vec{k})$ & 12  & 14 & 14 & 14 & 14 & 14 & 14 & 14
\end{tabular}
\caption{The number of Kramers pairs with $-1$ parity eigenvalues $n_{-}(\vec{k})$ at each TRIM point in $\beta$-MoTe$_2$ (SA~\ref{sec:DFTmethod}).  The $\mathbb{Z}_{4}$ index~\cite{ChenTCI,AshvinIndicators,AshvinTCI,EslamInversion} $\sum n_{-}(\vec{k})\text{ mod }4=2$.  Along with the trivial weak indices~\cite{FuKaneMele,FuKaneInversion,AdyWTI} and the absence of fourfold and sixfold rotation symmetries~\cite{WladTheory,HOTIBernevig,ChenRotation,HOTIChen,ChenTCI,AshvinTCI}, this indicates that $\beta$-MoTe$_2$ is a HOTI.}
\label{tb:z4}
\end{center}
\end{table}

We also explore the topology of the gapped regions between the MNLs by calculating the $z$-directed Wilson loop $W_{1}(k_{x},k_{y})$  (Fig.~\ref{fig:wloop}(a)) over the lower $N$ bands in the absence of SOC.  In BZ planes indexed by $k_{y}$ away from the MNLs, $W_{1}(k_{x},k_{y})$ exhibits gaps at $\theta_{1}/2\pi\approx \pm 0.25$ (Fig.~\ref{fig:wloop}(c,d) and SA~\ref{sec:DFTnested}), allowing us to calculate a nested Wilson loop matrix $W_{2}(k_{y})$ for which $\det(W_{2}(k_{y}))=\exp({i\gamma_{2}})$, where $\gamma_{2}$ is the nested Berry phase~\cite{multipole,WladTheory,HOTIBernevig}.  In bulk-gapped $k_{y}$-indexed planes, $\det(W_{2}(k_{y}))$ is quantized at $\pm 1$ (Fig.~\ref{fig:wloop}(b)), indicating that $\gamma_{2}= \pi$ ($0$) below (above) the MNL, implying that the Hamiltonians $\mathcal{H}_{k_{y}}(k_{x},k_{z})$ of planes in the two regions are topologically distinct.  This quantization can be understood from two perspectives: the bulk and the Wilson loop.  From a bulk perspective, $\mathcal{H}_{k_{y}}(k_{x},k_{z})$ is invariant under a local spinless time-reversal symmetry $\mathcal{I}\times\tilde{\mathcal{T}}$ that preserves the signs of $k_{x,z}$ and squares to $+1$.  $\mathcal{H}_{k_{y}}(k_{x},k_{z})$ therefore lies in Class AI of the Altland-Zirnbauer classification~\cite{AZClass,KitaevClass} with codimension~\cite{TeoKaneDefect,YoungkukLineNode} $D\mod 8 = 6$, implying a $\mathbb{Z}_{2}$ topology.  This topology can be diagnosed by considering the Wilson-loop perspective.  In SA~\ref{sec:DFTnested}, we show that $\mathcal{I}\times\bar{\mathcal{T}}$, which acts on $W_{1}(k_{x},k_{y})$ as an antiunitary particle-hole symmetry $\tilde{\Xi}$ that preserves the signs of $k_{x,y}$~\cite{ArisInversion,Cohomological,DiracInsulator}, enforces $\text{det}(W_{2}(k_{y}))=\pm 1$ when it is evaluated over any $\tilde{\Xi}$-symmetric grouping of Wilson bands (Fig.~\ref{fig:wloop}(c,d)), including (but not limited to) the Wilson bands between $\theta_{1}/2\pi\approx \pm 0.25$ in Fig.~\ref{fig:wloop}(c,d).  Crucially, building on~\onlinecite{multipole,WladTheory,HingeSM}, which employ a different definition of $\gamma_{2}$ reliant on fourfold rotation symmetry, the $\pi$ shift in $\gamma_{2}$ indicates that $\mathcal{H}_{k_{y}}(k_{x},k_{z})$ above and below an MNL are equivalent to topologically distinct 2D magnetic atomic limits~\cite{QuantumChemistry,HingeSM} (or trivialized fragile phases~\cite{AshvinFragile,JenFragile1,JenFragile2,HingeSM,AdrianFragile,WiederAxion}) that differ by the presence or absence of topological corner (hinge) modes (SA~\ref{sec:TBmodel}).  This implies that MNLSMs are \emph{higher-order} topological SMs~\cite{HingeSM,TaylorToy} with flat-band-like hinge states (Fig.~\ref{fig:doubleBand} and SA~\ref{sec:TBmodel}).  The jump in $\gamma_{2}$ as $k_{y}$ passes through an MNL (Fig.~\ref{fig:wloop}(b)) thus represents a new example of a topological ``descent relation,'' analogous to the jump in Berry phase as the line on which it is calculated passes through a Dirac point in 2D and an NL in 3D~\cite{YoungkukLineNode}.  Like in a Weyl SM~\cite{AG1985,NielsenNinomiya1}, the winding of the Wilson loop evaluated on a closed surface around the MNL (Fig.~\ref{fig:Fig2}(b)) captures the difference in topology between the gapped planes above and below it~\cite{YoungkukMonopole}, which, here, is the gapless point in $W_{1}$ (which is well defined when $W_{1}$ is evaluated on a slightly distorted path that avoids the MNLs).

\begin{figure}
\includegraphics[width=\linewidth]{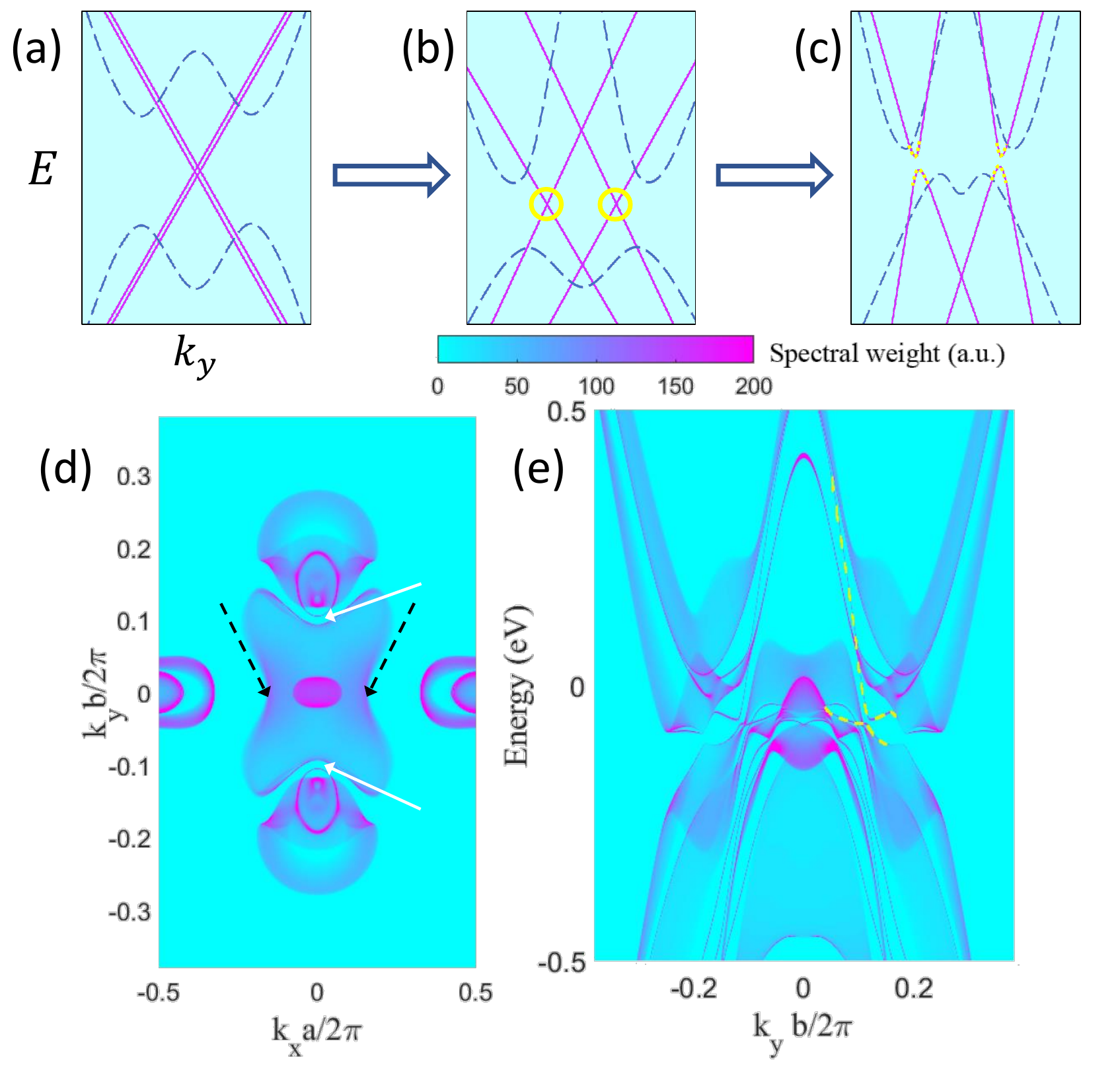}
\caption{(a-c) Schematic surface state evolution of a HOTI driven by DBI.  (a) Two bulk bands inverted at the same energy (blue dashed lines) realize a fourfold surface Dirac fermion (purple lines)~\cite{HOTIBismuth}.  (b) In the absence of specific glide reflection symmetries, this fermion is unstable~\cite{DiracInsulator}, and will split into two, twofold surface fermions, which may be stabilized by either a surface mirror (topological crystalline insulator (TCI))~\cite{TeoFuKaneTCI,HsiehTCI}, glide (hourglass TCI)~\cite{HourglassInsulator,DiracInsulator}, or $C_{2z}\times\mathcal{T}$ symmetry (rotation anomaly TCI)~\cite{ChenRotation,FangFuNSandC2,WiederAxion}.  (c) In the absence of surface reflection or rotation symmetries, the twofold cones (yellow circles in (b), dashed lines in (c)) hybridize and gap, realizing the surface of a HOTI~\cite{HOTIBernevig}.  (d) Spectral weight at $E_{F}$ of states on the $(001)$ surface of $\beta$-MoTe$_2$ plotted as a function of $k_{x,y}$, and (e) along $k_{x}=0$ as a function of energy (SA~\ref{sec:DFTmethod}).  Each of the two band inversions at the bulk $\Gamma$ point (Fig.~\ref{fig:Fig2}(d)) nucleates a topological twofold surface cone centered at $k_{x}=k_{y}=0$ (purple); the cones then repel each other in energy and merge with the projections of the bulk states (b).  As depicted in (c), the surface bands from these cones (white arrows in (d)) hybridize and gap (yellow dashed lines in (e)) to form a narrowly avoided crossing.  In $\gamma$-XTe$_{2}$, these hybridized cones also appear as surface states~\cite{wang_mote2:_2016,soluyanov_type-ii_2015,sun_prediction_2015,WTe2Arpes1,WTe2Arpes2,WTe2Arpes3,WTe2Arpes4,MoTe2Arpes1,MoTe2Arpes2,MoTe2Arpes3,MoTe2Arpes4,WTe2STM,WTe2QPI2018,MoTe2STM}, but their gap is spanned along $k_{x}$ by small, topological Fermi arcs from bulk type-II Weyl points below $E_{F}$ (black arrows in (d)).  In SA~\ref{sec:DFTgappedGamma}, we calculate the $(001)$ surface states of $\gamma$-MoTe$_{2}$ gapped with symmetry-preserving distortion, and observe gapped HOTI surface states nearly identical to those in $\beta$-MoTe$_2$ (d,e).}
\label{fig:surf}
\end{figure}

When SOC is introduced, $\beta$-MoTe$_2$, though remaining metallic, develops a gap at all crystal momenta (Fig.~\ref{fig:Fig2}(d)).  Calculating the parity eigenvalues (Table~\ref{tb:z4}), we find that, though the Fu-Kane $\mathbb{Z}_{2}$ index~\cite{FuKaneMele,FuKaneInversion} is trivial, the occupied bands nevertheless cannot be expressed as a sum of EBRs, indicating an overall nontrivial topology~\cite{QuantumChemistry}.  Specifically, while every EBR in an $\mathcal{I}$-symmetric space group exhibits Kramers pairs of parity eigenvalues~\cite{QuantumChemistry,Bandrep1} for which $\sum n_{-}(\vec{k})\text{ mod }4=0$, the DBI in $\beta$-MoTe$_{2}$ induces an insulator with $\sum n_{-}(\vec{k})\text{ mod }4 = 2$ (Table~\ref{tb:z4}).  Alternatively, this defines a $\mathbb{Z}_{4}$ index~\cite{ChenTCI,AshvinIndicators,AshvinTCI,EslamInversion} which is nontrivial.  From both perspectives, $\beta$-MoTe$_2$ carries the parity eigenvalues of a HOTI~\cite{HOTIBernevig,HOTIBismuth}.  Therefore, like bismuth~\cite{BismuthLayer1,BismuthLayer2,HOTIBismuth}, $\beta$-MoTe$_2$ is a 2D TI when viewed as a quasi-2D system~\cite{LiangTMD,TMDTI,WTe2TI}, but is actually a HOTI when taken to be fully 3D.

Unlike $\beta$-MoTe$_{2}$, $\beta$-WTe$_2$, while stabilizable as a monolayer~\cite{CarlSynthesis2,WTe2TI}, is unstable as a bulk crystal~\cite{UnstableBetaWTe2}.  Nevertheless, the calculated electronic structure of artificial $\beta$-WTe$_2$ also exhibits DBI at the $\Gamma$ point~\cite{UnstableBetaWTe2}, indicating that it would also be a HOTI if it could be stabilized.  However, we will shortly see that remnants of this HOTI phase are still observable in $\gamma$-WTe$_2$.

In Fig.~\ref{fig:surf}(d,e), we plot the $(001)$ surface states of $\beta$-MoTe$_2$ calculated from first-principles (SA~\ref{sec:DFTmethod}).  We observe large, arc-like surface states around the projection of the $\Gamma$ point (white arrows in (d)), as well the projections of bulk states at the $k_{y}=0$ surface TRIM points.  Crucially, we determine that the surface states are in fact gapped at all values of $k_{x,y}$.  This can be understood by considering the symmetry and topology consequences of the bulk DBI at $\Gamma$.  In the absence of SOC, each band inversion nucleates one drumhead surface state (per spin) around the surface projection of the $\Gamma$ point (SA~\ref{sec:TBmodel}).  In the absence of additional surface wallpaper group symmetries, such as mirror or glide~\cite{DiracInsulator,WiederLayers}, these drumhead states hybridize and gap.  When SOC is reintroduced, the four hybridized drumhead states (two per spin) open into two hybridized twofold surface TI cones (Fig.~\ref{fig:surf}(a-c)).  Therefore rather than being trivial Fermi arcs, the surface states of $\beta$-MoTe$_2$ are in fact the characteristic split and gapped fourfold Dirac-cone states of a HOTI~\cite{HOTIBernevig,HOTIBismuth,DiracInsulator}.  Unlike the gapless surface states of 3D TIs~\cite{FuKaneMele,FuKaneInversion}, the gapped Fermi arcs only appear at low energies because of the interplay of SOC and band dispersion in $\beta$-MoTe$_2$, and could theoretically be moved away from $E_{F}$ without changing the bulk or surface topology (SA~\ref{sec:TBmodel}).

This observation solves a longstanding mystery in $\gamma$-XTe$_{2}$.  In theoretical predictions~\cite{wang_mote2:_2016,soluyanov_type-ii_2015,sun_prediction_2015} and bulk experimental probes~\cite{BulkMoTe2,BulkWTe21,BulkWTe22}, both $\gamma$-MoTe$_2$ and $\gamma$-WTe$_2$ exhibit narrowly separated type-II Weyl points in the vicinity of doubly inverted bands.  Nevertheless, as measured both directly by ARPES~\cite{WTe2Arpes1,WTe2Arpes2,WTe2Arpes3,MoTe2Arpes1,MoTe2Arpes2} and through quasiparticle interference in STM probes~\cite{WTe2STM,WTe2QPI2018,MoTe2STM}, $\gamma$-XTe$_2$ crystals also exhibit huge, arc-like surface states that largely overwhelm possible signatures of topological Weyl Fermi arcs.  Previous works determined these large surface arcs to be topologically trivial~\cite{wang_mote2:_2016,soluyanov_type-ii_2015,sun_prediction_2015,WTe2Arpes2,WTe2Arpes3,WTe2Arpes4,MoTe2Arpes1,MoTe2Arpes2,MoTe2Arpes4,WTe2STM,WTe2QPI2018,MoTe2STM}.  However, in light of our previous analysis of similar \emph{nontrivial} surface states in $\beta$-MoTe$_2$, we recognize this determination to be incomplete.  By explicitly calculating the surface states and bulk topology of $\gamma$-MoTe$_2$ when its Weyl points are gapped by slight distortion (SA~\ref{sec:DFTgappedGamma}), we instead discover that the large Fermi arcs in $\gamma$-XTe$_2$ represent the split surface Dirac cones of a \emph{non-symmetry-indicated} HOTI phase that is nearby in parameter space and driven by DBI (Fig.~\ref{fig:surf}).  Given the small separation of the bulk Weyl points, which is quite sensitive to experimental conditions~\cite{WTe2Arpes1,MoTe2STM,gapTrans1,gapTrans2}, this HOTI phase may be accessible via symmetry-preserving distortion or strain, and may already be realized in existing samples.

In this letter, we have demonstrated higher-order topology in both $\beta$- and $\gamma$-XTe$_2$.  When SOC is neglected in $\beta$-MoTe$_2$, we observe a pair of MNLs at $E_{F}$.  While SOC cannot be neglected in $\beta$-MoTe$_2$, other centrosymmetric materials with lighter atoms and DBI are likely to also exhibit MNLs~\cite{YoungkukMonopole}.  When the effects of SOC are incorporated, $\beta$-MoTe$_2$ develops a direct gap with the parity eigenvalues of a HOTI.  Though $\beta$-MoTe$_2$ is in fact metallic, and thus does not host the projected hinge gap required to observe its characteristic helical hinge modes~\cite{HOTIBismuth}, it is possible that a TMD with more favorable band dispersion could be engineered by intercalation or chemical substitution.  Finally, we observe that both $\gamma$-MoTe$_2$ and $\gamma$-WTe$_2$ exhibit the same large topological surface arcs as $\beta$-MoTe$_2$, resolving an outstanding puzzle in TMDs, and presenting a new venue for investigating non-symmetry-indicated higher-order topology.

\begin{acknowledgements}
The authors thank Barry Bradlyn, Jennifer Cano, Yichen Hu, Fan Zhang, and Youngkuk Kim for helpful discussions.  Z. W., B. J. W., and B. A. B. were supported by the Department of Energy Grant No. DE-SC0016239, the National Science Foundation EAGER Grant No. DMR 1643312 and NSF-MRSEC No. DMR-1420541, Army Research Office Grant No. ARO MURI W911NF-12-1-0461, Simons Investigator Grant No. 404513, ONR Grant No. N00014-14-1-0330, the Packard Foundation, the Schmidt Fund for Innovative Research, and a Guggenheim Fellowship from the John Simon Guggenheim Memorial Foundation.   Z. W. additionally acknowledges support from the National Thousand-Young-Talents Program and the CAS Pioneer Hundred Talents Program. This work was also supported by the National Natural Science Foundation of China (No. 11504117 and No. 11774317).  B.Y. acknowledges support from the Willner Family Leadership Institute for the Weizmann Institute of Science, the Benoziyo Endowment Fund for the Advancement of Science, the Ruth and Herman Albert Scholars Program for New Scientists, and the European Research Council (ERC) under the European Union Horizon 2020 Research and Innovation Programme (Grant No. 815869).  During the final stages of preparing this manuscript, hinge states in $\beta$-MoTe$_2$ were also predicted in~\onlinecite{AshvinTCIMaterials}, but the relationship between higher-order topology in $\beta$-MoTe$_2$, MNLs, Fermi arcs in $\gamma$-XTe$_2$, and the experimental data was not previously established.  After the submission of this letter, we became aware that, in a revised version of~\onlinecite{YoungkukMonopole}, fragile topology was also recognized in 2D insulators with $\mathcal{I}\times\tilde{\mathcal{T}}$ symmetry.  The fragile-phase corner charges introduced in this letter were subsequently verified and further explored in~\onlinecite{KoreanFragile,WiederAxion,KoreanAxion,WladCorners,KoreanInversionFragile}, and were demonstrated in~\onlinecite{WiederAxion,KoreanAxion,BJYangVortex}, along with the alternative formulation of the nested Wilson loop introduced in this letter (SA~\ref{sec:DFTnested}), to be essential components of the pumping formulation of axion insulators.  The flat-band-like MNL hinge states predicted in this letter were subsequently observed in first-principles calculations of 3D graphdiyne~\cite{GraphdiyneHinge}.  The nested Jackiw-Rebbi formulation of fragile-phase corner charges employed in this letter (SA~\ref{sec:TBmodel}, adapted from~\onlinecite{HingeSM}, which was researched concurrently with this letter) was subsequently generalized in~\onlinecite{KoreanInversionFragile} to characterize the (anomalous) 0D boundary modes of $\mathcal{I}$-symmetric fragile phases with arbitrary dimensionality.  Finally, after the submission of this letter, incipient experimental signatures of hinge states in MoTe$_2$ and WTe$_2$ were observed in~\onlinecite{DavidMoTe2,MazHingeExp}, respectively.
\end{acknowledgements}

\clearpage
\onecolumngrid
\begin{appendix}

\begin{center}
{\bf Supplementary Appendices for ``Higher-Order Topology, Monopole Nodal Lines, and the Origin of Large Fermi Arcs in Transition Metal Dichalcogenides XTe$_2$ (X$=$Mo,W)''}
\end{center}

\section{Tight-Binding Model for Flat-Band Hinge States in a Monopole Nodal-Line Semimetal}
\label{sec:TBmodel}

\begin{figure}[h]
\centering
\includegraphics[width=0.90\textwidth]{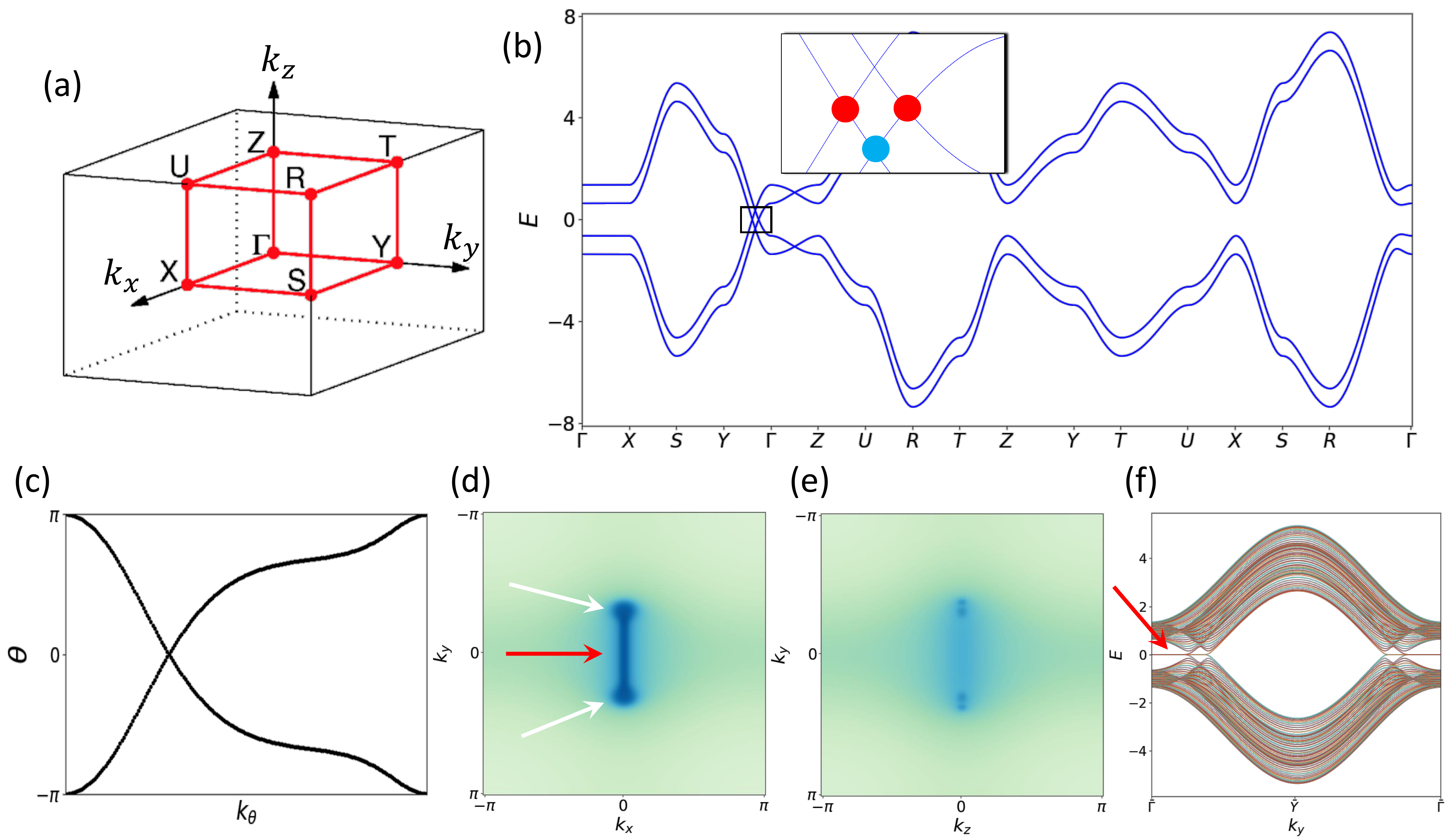}
\caption{(a) Primitive orthorhombic BZ~\cite{BCTBZ}.  (b) Bulk bands for $\tilde{\mathcal{H}}(\vec{k})$ (Eq.~(\ref{eq:simpleHam})) with the parameters in Eq.~(\ref{eq:uncoupledParams}).  In this limit, the eight-band model exhibits an overall $SU(2)$ spin rotation symmetry, and bands therefore appear in spin-degenerate pairs.  In the inset panel, we highlight a monopole nodal line (MNL) at $E_{F}=0$ along $Y\Gamma$ (red circles) that is linked to its time-reversal partner by nodal lines encircling the $\Gamma$ point directly above and below $E_{F}$ (blue circle) (axes not in parenthesis in Fig.~\ref{fig:coupled}(d)), as discussed in Ref.~\onlinecite{YoungkukMonopole}.  Using the bulk tight-binding model (Eq.~(\ref{eq:simpleHam})), we deduce that the MNLs lie in the $k_{xy}$-plane.  (c) The Wilson loop eigenvalues over the lower four occupied bands calculated on a sphere surrounding one of the MNLs, plotted as a function of the polar momentum $k_{\theta}$ (Fig.~\ref{fig:Fig2}(c) of the main text), exhibit helical winding.  As detailed in Refs.~\onlinecite{YoungkukMonopole,AdrianMonopole,SigristMonopole}, the helical winding in (c) confirms that the nodal line enclosed by the sphere carries a nontrivial monopole charge,  and thus, is an MNL.  (d,e) The $(001)$ and $(100)$ surface states of Eq.~(\ref{eq:simpleHam}), respectively, calculated at E=0.  In (d), we observe both topological drumhead states on the interior projections of the bulk MNLs (white arrows)~\cite{YoungkukLineNode} and extraneous surface states (red arrows) that are remnants of the bulk double band inversion.  Specifically, the interior states indicated by the white arrows are topologically protected~\cite{YoungkukMonopole}, whereas the line of states shown by the red arrows is topologically trivial, and lies outside of the projections of the MNLs.  The extraneous states originate from the double band inversion shown in Fig.~\ref{fig:DBI}: the first band inversion creates a drumhead state at $k_{x}=k_{y}=0$ of the $(001)$ surface BZ, and the pinching process~\cite{YoungkukMonopole} of the nodal line at half filling forms a second set of surface states, rather than removing the drumhead states from the first band inversion.  The extraneous surface states are topologically trivial, and represent an artificial degeneracy in the limit of the parameters in Eq.~(\ref{eq:uncoupledParams}); we add the terms necessary to hybridize and gap the extra surface states in Fig.~\ref{fig:coupled}.  (f) The bands of a $z$-directed slab of the model in (b), plotted at $k_{x}=0$ as a function of $k_{y}$.  The extraneous surface states from (d) are also marked in (f) with a red arrow.}
\label{fig:uncoupled}
\end{figure}

In this section, we construct a model of a 3D nodal-line semimetal (NLSM)~\cite{YoungkukLineNode,XiLineNode,SchnyderDrumhead} that exhibits a time-reversed pair of nodal lines with nontrivial monopole charges (MNLs)~\cite{FangWithWithout} generated by double band inversion~\cite{YoungkukMonopole}.  Using this model, we demonstrate that, when the appropriate coupling terms are added, bulk and surface gaps generically develop in the momentum-space region between the two MNLs (Fig.~\ref{fig:coupled}(d)).  Projecting the bulk MNLs and surface drumhead states to the 1D hinges, additional 1D, flat-band-like surface states can be observed spanning the region between the projections of the MNLs.  These states represent the $d-2$-dimensional generalization of drumhead surface states, and are the spinless analogs of the flat-band-like hinge states recently predicted in spinful Dirac semimetals~\cite{HingeSM,TaylorToy}.  In this section, we show that this monopole NLSM (MNLSM) can be gapped into either a spinless magnetic higher-order topological insulator (HOTI) (otherwise known as an ``axion insulator''~\cite{EzawaMagneticHOTI,VDBAxion,YoungkukMonopole,VDBHOTI,WiederAxion,KoreanAxion}), or into a spinful, time-reversal- ($\mathcal{T}$-) symmetric HOTI with helical pairs of hinge modes.  All calculations for this section were performed employing the~\textsc{PythTB} package~\cite{PythTB}.

To begin, we place eight spinful orbitals at the origin of a primitive orthorhombic unit cell; these can be considered four spinless orbitals (two $s$ and two $ip$ orbitals), each with an additional (initially uncoupled) spin-$1/2$ degree of freedom.  We index the four pairs of spinful orbitals with the Pauli matrices $\tau$ and $\mu$, and index the spin degree of freedom with $\sigma$.  Using these orbitals, we construct the tight-binding Hamiltonian:
\begin{equation}
\tilde{\mathcal{H}}(\vec{k}) = \left[m_{1} + \sum_{i=x,y,z} v_{i} \cos(k_{i})\right]\tau^{z} + m_{2}\tau^{z}\mu^{x} + m_{3}\tau^{z}\mu^{z} + u_{x}\sin(k_{x})\tau^{y}\mu^{y} + u_{z}\sin(k_{z})\tau^{x},
\label{eq:simpleHam}
\end{equation}
which is invariant under inversion ($\mathcal{I}$) and spinless time-reversal ($\tilde{\mathcal{T}}$), represented by their action on the Hamiltonian as:
\begin{equation}
\tilde{\mathcal{T}}\mathcal{H}(\vec{k})\tilde{\mathcal{T}}^{-1} = \tau^{z}\mathcal{H}^{*}(-\vec{k})\tau^{z},\ \mathcal{I}\mathcal{H}(\vec{k})\mathcal{I}^{-1} = \tau^{z}\mathcal{H}(-\vec{k})\tau^{z},
\label{eq:symmetries}
\end{equation}
as well as, at first, $SU(2)$ spin symmetry.  There are also other, extraneous symmetries, which we are free to break.  For simplicity, we have chosen units in which the lattice constants $a_{x,y,z}=1$.  The representations of the symmetries in Eq.~(\ref{eq:symmetries}) are chosen in a basis ($s$ and $ip$ orbitals at $(x,y,z)=(0,0,0)$) for which the combined antiunitary symmetry $\mathcal{I}\times\tilde{\mathcal{T}}=K$ guarantees that all of the $4\times 4$ matrix coefficients of Eq.~(\ref{eq:simpleHam}) are real~\cite{FangWithWithout,YoungkukMonopole}.  The form of $\tilde{\mathcal{H}}(\vec{k})$ is specifically chosen to include terms from the models in Refs.~\onlinecite{FangWithWithout,YoungkukMonopole}, as to guarantee that $\tilde{\mathcal{H}}(\vec{k})$ will form a pair of MNLs after undergoing double band inversion about the $\Gamma$ point.

\begin{figure}[h]
\centering
\includegraphics[width=\textwidth]{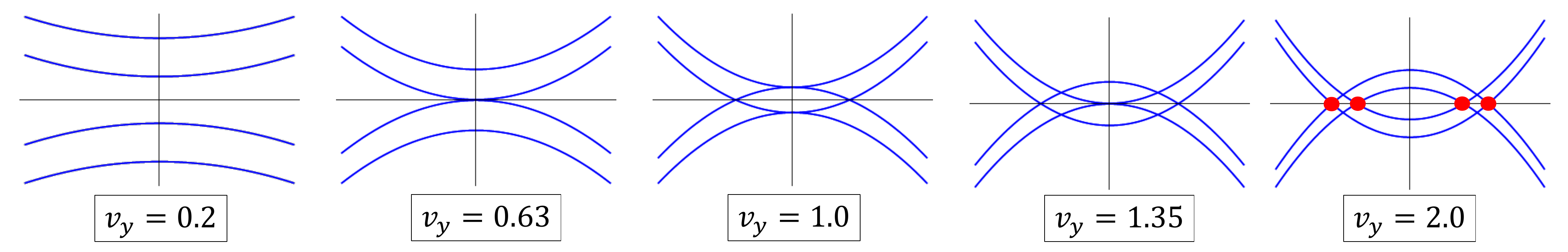}
\caption{Bulk bands of $\tilde{\mathcal{H}}_{C}(\vec{k})$ in Eq.~(\ref{eq:coupledHam}), plotted along $Y\Gamma$ in the vicinity of the $\Gamma$ point with the parameters in Eqs.~(\ref{eq:uncoupledParams}) and varying $v_{y}$.  As $v_{y}$ is tuned between $0$ and $2$, bands become doubly inverted, and eventually form a time-reversed pair of MNLs at half filling (red circles at $v_{y}=2.0$), as described in Ref.~\onlinecite{YoungkukMonopole}.}
\label{fig:DBI}
\end{figure}

We track the bulk phase transitions of Eq.~(\ref{eq:simpleHam}) by choosing the parameters:
\begin{equation}
m_{1}=-3,\ v_{x}=v_{z}=u_{x}=u_{z}=1,\ m_{2}=0.3,\ m_{3}=0.2.
\label{eq:uncoupledParams}
\end{equation}
and tuning $v_{y}$ between $0$ and $2$ (Fig.~\ref{fig:DBI}).  When $v_{y}\approx 0.63$, bands begin to invert at $\Gamma$, forming a nodal line (without monopole charge) between the second and third pair of spin-degenerate bands.  When $v_{y}$ reaches $1$, the third pair of bands reaches the first pair of bands (and the second touches the fourth), and a nodal line at quarter filling (shown in blue in Fig.~\ref{fig:coupled}(d)) begins to form.  Next, when $v_{y}\approx 1.35$, the nodal line at half filling begins to pinch off into a time-reversed pair of nodal lines that intersect $Y \Gamma$ (red lines in Fig.~\ref{fig:coupled}(d)).  We then finally tune $v_{y}\rightarrow 2$ to grow the two nodal lines at half filling (red circles in Fig.~\ref{fig:DBI}) to have clearly distinguishable interior regions.  Calculating the $(001)$ and $(100)$ surface Green's function of Eq.~(\ref{eq:simpleHam}) at $E_{F}=0$ (Fig.~\ref{fig:uncoupled}(d,e)), we observe the presence of drumhead surface states (white arrows) on only the $z$-normal ($(001)$) surface, indicating that the bulk nodal lines are almost entirely normal to the $k_{z}$-axis (Fig.~\ref{fig:coupled}(d), axes not in parenthesis).  To calculate the $\mathbb{Z}_{2}$ monopole charge of each nodal line, we surround it with an approximate sphere and calculate the Wilson loop eigenvalues over the lower four bands (including spin) as a function of the polar momentum ($k_{\theta}$ in Fig.~\ref{fig:Fig2}(c) of the main text), as prescribed in Refs.~\onlinecite{YoungkukMonopole,AdrianMonopole,SigristMonopole}.  Most precisely, we approximate this sphere by calculating the Wilson loop on a series of concentric $k_{y}$-normal circles, indexed by $k_{y}$, and centered at $k_{x}=k_{z}=0$, and where the radii of the circles are tapered above and below the nodal line at half filling.  The Wilson loop spectrum (Figs.~\ref{fig:uncoupled}(c)) exhibits clear helical winding, confirming the nontrivial monopole charge of each nodal line at half filling~\cite{YoungkukMonopole,AdrianMonopole,SigristMonopole}.

This helical winding can be understood as reconciling the topology of the gapped planes indexed by $k_{y}$ above and below the MNL.  As shown in Ref.~\onlinecite{YoungkukMonopole}, the Hamiltonians of the two gapped planes are equivalent to topologically distinct 2D insulators, and can be distinguished by their (gapped) Wilson spectra.  As the Wilson loop on a sphere can be deformed into the Wilson loop on the plane above the sphere minus the Wilson loop on the plane below the sphere (the prototypical explanation for the conservation of Chern number in a Weyl semimetal~\cite{AG1985,NielsenNinomiya1}), we recognize that the gapless Wilson spectrum on the sphere reflects the Wilson loop critical point that distinguishes the 2D insulating phases above and below the MNL~\cite{YoungkukMonopole}.  More precisely, Ref.~\onlinecite{YoungkukMonopole} establishes a $\mathbb{Z}_{2}\times\mathbb{Z}_{2}$ classification of the possible topologies of the Hamiltonians of the $\mathcal{I}\times\tilde{\mathcal{T}}$-symmetric planes above and below an MNL, and the helical winding of the sphere Wilson loop indicates a change in one of these indexes.  To understand the topology of the planes with only $\mathcal{I}\times\tilde{\mathcal{T}}$ symmetry, we will first consider a topologically nontrivial 2D Hamiltonian with both $\mathcal{I}$ and $\tilde{\mathcal{T}}$ symmetry ($k_{y}=0$), and will then subsequently break those symmetries in a manner that preserves their product $\mathcal{I}\times\tilde{\mathcal{T}}$ and does not close a bulk or edge (surface) gap.

\begin{figure}[h]
\centering
\includegraphics[width=0.6\textwidth]{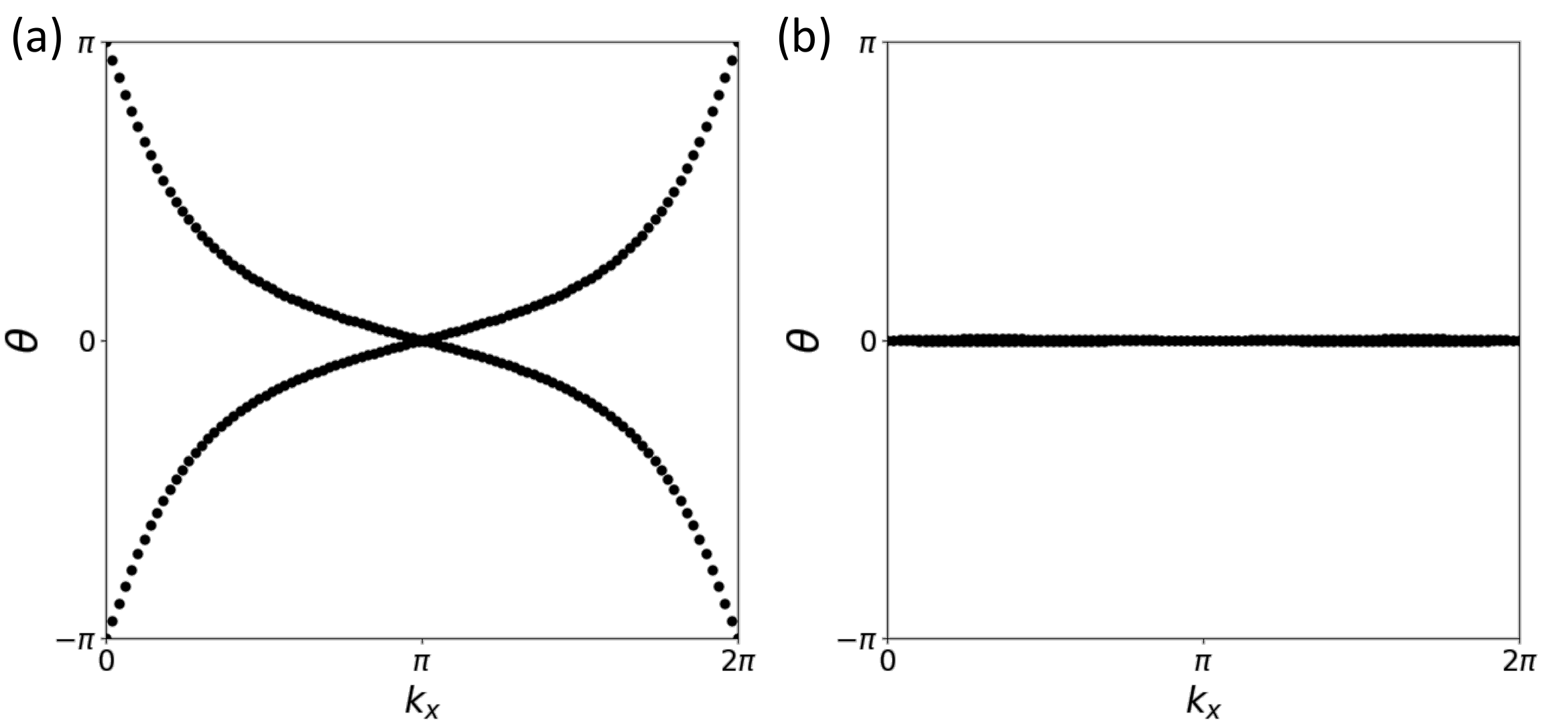}
\caption{(a,b) $z$-directed Wilson loops evaluated for the lowest two spinless pairs of bands in Fig.~\ref{fig:uncoupled}(a) at $k_{y} = 0,\pi$, respectively.  The winding of the Wilson loop at $k_{y}=0$ is not protected by spinful time-reversal symmetry, as it would be in a 2D topological insulator~\cite{AndreiTI}, but is instead protected by the combination of the bulk inversion eigenvalues~\cite{ArisInversion} and the absence of additional bands in the Wilson projector, and thus is an example of ``fragile'' topology~\cite{AshvinFragile,JenFragile1,JenFragile2,HingeSM,AdrianFragile,WiederAxion}.  At $k_{y}=\pi$ (b), the bulk inversion eigenvalues require that $\theta = 0$ at $k_{y}=0,\pi$~\cite{ArisInversion}.  We observe that the $z$-directed Wilson loop in the $k_{y}=\pi$ plane exhibits trivial winding, and extremely weak dispersion.}
\label{fig:planeWilson}
\end{figure}

Using the results of Ref.~\onlinecite{ArisInversion}, we determine that the Hamiltonian of the $k_{y}=0$ plane exhibits a gapless $z$-directed Wilson loop with nontrivial winding (Fig.~\ref{fig:planeWilson}(a)), because all of the occupied bands have the same inversion eigenvalues at each TRIM point, and because the inversion eigenvalues at the $\Gamma$ point differ from those at the three other 2D TRIM points.  The winding of this Wilson loop may be removed by adding trivial bands with different inversion eigenvalues~\cite{ArisInversion} (as occurs in the $k_{y}=0$ plane of $\beta$-MoTe$_2$ (Fig.~\ref{fig:wloop}(c) of the main text)); this is a hallmark of a fragile topological phase~\cite{AshvinFragile,JenFragile1,JenFragile2,HingeSM,AdrianFragile,WiederAxion}.  Unlike in a 2D TI, we will show that the fragile Wilson loop winding at $k_{y}=0$ indicates the presence of nontrivial \emph{corner} modes.  Specifically, using the $k\cdot p$ theory of Eq.~(\ref{eq:simpleHam}), we will exploit the nested Jackiw-Rebbi construction from Ref.~\onlinecite{HingeSM}, which was developed concurrently with this letter, to demonstrate that the Hamiltonian of the $k_{y}=0$ plane is equivalent to a 2D insulator with gapped edges and spin-degenerate pairs of corner modes.  First, we expand Eq.~(\ref{eq:simpleHam}) about the $\Gamma$ point in the limit that $m_{2}=m_{3}=0$, $u_{x}=u_{z}=u$:
\begin{equation}
\mathcal{H}_{\Gamma}(\vec{k}) = m\tau^{z} + u(\tau^{y}\mu^{y}k_{x} + \tau^{x}k_{z}),
\end{equation}
where we have condensed all of the terms proportional to $\tau^{z}$ into a single mass term $m$.  We then take $m$ to vary spatially, such that it is negative in a circular region bound by a radius $R$ and large and positive outside of it.  The bound state solutions on the exterior of this region can be obtained by forming a Jackiw-Rebbi domain wall~\cite{JackiwRebbi} at $r=R$, which we accomplish by Fourier transforming $k_{x,z}\rightarrow -i\partial_{x,z}$ and converting to polar coordinates:
\begin{equation}
\mathcal{H}(r,\theta) = m(r)\tau^{z} - iu\Gamma^{1}(\theta)\partial_{r} - \frac{iu}{r}\Gamma^{2}(\theta)\partial_{\theta},
\label{eq:mainHamiltonianCircle}
\end{equation}
where:
\begin{equation}
\Gamma^{1}(\theta) = \tau^{x}\cos(\theta) + \tau^{y}\mu^{y}\sin(\theta),\ \Gamma^{2}(\theta)=-\tau^{x}\sin(\theta) + \tau^{y}\mu^{y}\cos(\theta).
\end{equation}
In the absence of additional terms, $\mathcal{H}(r,\theta)$ exhibits gapless, linear dispersing modes on its edges, as in this limit it is closely related to the $k\cdot p$ theory of a 2D topological insulator~\cite{AndreiTI,HingeSM}.  More specifically, despite having only spinless time-reversal symmetry, $\mathcal{H}(r,\theta)$ still exhibits linear dispersion with three anticommuting $4 \times 4$ Dirac matrices, like the $k\cdot p$ theory of the Bernevig-Hughes-Zhang model of a 2D TI~\cite{AndreiTI,HingeSM}.  In polar coordinates, the symmetries of $\mathcal{H}(r,\theta)$ are represented by their action: 
\begin{equation}
\tilde{\mathcal{T}}\mathcal{H}(r,\theta)\tilde{\mathcal{T}}^{-1} = \tau^{z}\mathcal{H}^{*}(r,\theta)\tau^{z},\ \mathcal{I}\mathcal{H}(r,\theta)\mathcal{I}^{-1} = \tau^{z}\mathcal{H}(r,\theta+\pi)\tau^{z}.
\label{eq:symmetriesPolar}
\end{equation}
The symmetries in Eq.~(\ref{eq:symmetriesPolar}) permit a set of bulk mass terms that includes:
\begin{equation}
V_{L}(\theta) = \tau^{y}\mu^{x}\sin(L\theta + \phi),\ L=L^{FTI} = 1 + 2n,\ n\in\mathbb{Z},
\label{eq:nestedJackiw}
\end{equation}
where $V_{L}(\theta)$ anticommutes with all of the existing terms in $\mathcal{H}(r,\theta)$, and is therefore guaranteed to open bulk (and edge) gaps in all of the regions in which it is nonzero.  We use the label ``$FTI$'' on $L$ in Eq.~(\ref{eq:nestedJackiw}) to emphasize that $V_{L}(\theta)$ is the intrinsic bulk mass of a fragile TI.  In Eq.~(\ref{eq:nestedJackiw}), the integer $L$ is the ``angular momentum'' of the mass term~\cite{HingeSM}, and $\phi$ is a free angle.  When $L$ takes its lowest symmetry-allowed value ($L=L^{FTI}=1$), $V_{L}(\theta)$ becomes proportional to the circular harmonic~\cite{HingeSM} of a $p$ orbital lying in the $xz$-plane whose lobes are oriented at an angle $\phi$ from the $x$ axis.  The Hamiltonian $\mathcal{H}(r,\theta) + V_{L}(\theta)$ therefore exhibits $2L = 2 + 4n$ spin-degenerate pairs of 0D bound states, where each pair of bound states at $\theta$ is related by $\mathcal{I}$-symmetry to a second pair at $\theta+\pi$ (see Ref.~\onlinecite{HingeSM} for a more explicit derivation of the form of these 0D states and the role of curvature in this geometry).  Other mass terms are also allowed; however, in the $4\times 4$ basis of $\tau^{i}\otimes\mu^{i}$, all $\mathcal{I}$-symmetric bulk mass terms that anticommute with $\tau^{z}$ will necessarily also carry a spatial distribution $\sin(L^{FTI}\theta +\phi)$.  This indicates that the number of spin-degenerate pairs of 0D boundary modes modulo $4$ is an intrinsic property of $\mathcal{H}(r,\theta)$, with the smallest number (and the number seen in our numerics (Fig.~\ref{fig:coupled}(f)) being $2$.  We therefore conclude that the $k_{y}=0$ plane of Eq.~(\ref{eq:simpleHam}) is an additional example of a ``fragile'' topological phase~\cite{AshvinFragile,JenFragile1,JenFragile2,AdrianFragile,HingeSM,WiederAxion} that exhibits anomalous corner modes~\cite{HingeSM,WiederAxion} on the boundary of a finite-sized region with inversion symmetry.  Furthermore, at half-filling, each pair of 0D states exhibits a charge $\pm e/2$ per spin~\cite{WiederAxion,HingeSM}, where, specifically, $\mathcal{I}$-related bound states exhibit opposite charges~\cite{WiederAxion} within each spin sector indexed by $\sigma^{z}$.  As discussed in Ref.~\onlinecite{HingeSM}, this can be reformulated as the statement that the boundary of $\mathcal{H}(r,\theta) + V_{L}(\theta)$ exhibits a set of 0D charges with a total dipole moment per spin given by the sum of $L^{FTI}=1+2n$ free-angle dipoles, where each dipole has the same magnitude of $e/2$ per unit length .   Therefore, as first noted in Ref.~\onlinecite{HingeSM}, Eqs.~(\ref{eq:mainHamiltonianCircle}) and~(\ref{eq:nestedJackiw}) bear similarities with recent gauge-theory descriptions of fractons with anomalous ``vector'' charges~\cite{Fracton}.

We can model the process of moving to nearby planes indexed by $k_{y}\neq 0$ by considering the effect of introducing a bulk (spinless) mass term to Eqs.~(\ref{eq:mainHamiltonianCircle}) and~(\ref{eq:nestedJackiw}) that breaks $\mathcal{I}$ and $\tilde{\mathcal{T}}$ while preserving their product.  As we are still preserving $SU(2)$ spin symmetry, any additional term that does not close the bulk or edge gaps can only act as an $\mathcal{I}\times\tilde{\mathcal{T}}$-symmetric chemical potential on the corner modes~\cite{HingeSM}.  As previously shown, the two midgap corner modes at $k_{y}=0$ of the $L=1$ term in Eq.~(\ref{eq:nestedJackiw}) are related by $\mathcal{I}$ symmetry, and therefore remain anomalous if $\mathcal{I}$ is relaxed while preserving the combined operation $\mathcal{I}\times\tilde{\mathcal{T}}$.  When extra (trivial) bands without corner modes are added to remove the Wilson-loop winding, the $k_{y}=0$ plane of Eq.~(\ref{eq:simpleHam}) should exhibit a gapped Wilson spectrum with the same $\mathbb{Z}_{2}$-quantized nested Berry phase as the $k_{y}=0$ plane of $\beta$-MoTe$_2$ (Fig.~\ref{fig:wloop}(c) of the main text and Appendix~\ref{sec:DFTnested}, confirmed by explicit calculation in Ref.~\onlinecite{WiederAxion} after the submission of this letter).  Therefore, when a bulk term is added that preserves $\mathcal{I}\times\tilde{\mathcal{T}}$ while breaking the individual symmetries $\mathcal{I}$ and $\tilde{\mathcal{T}}$, the nested Berry phase will \emph{remain} quantized and the corner modes will remain present.  This indicates that, as all of the Hamiltonians of the $k_{y}$-indexed planes between the MNLs in our tight-binding model can be connected to the Hamiltonian of the plane at $k_{y}=0$ without closing a bulk or edge gap, they should also carry corner (hinge) modes indicated by a quantized nested Berry phase $\gamma_{2}=\pi$.  Utilizing the Wannier description of corner-mode phases developed in Ref.~\onlinecite{HingeSM}, this suggests that the four possible insulating phases of the Hamiltonians of the $\mathcal{I}\times\tilde{\mathcal{T}}$-symmetric $k_{y}\neq0$ planes between the MNLs in Eq.~(\ref{eq:simpleHam}), which were determined in Ref.~\onlinecite{YoungkukMonopole} to have a $\mathbb{Z}_{2}\times\mathbb{Z}_{2}$ topological classification, correspond to either \emph{non-symmetry-indicated} obstructed atomic limits~\cite{QuantumChemistry} or trivialized fragile topological insulators (for a large number of occupied bands)~\cite{AshvinFragile,JenFragile1,JenFragile2,AdrianFragile,HingeSM,WiederAxion} that differ by the number of Wannier orbitals on each of the four $\mathcal{I}\times\tilde{\mathcal{T}}$ centers of magnetic layer group $p\bar{1}'$~\cite{MagneticBook,HingeSM}.  After the submission of this letter, this Wannier description of $\mathcal{I}\times\tilde{\mathcal{T}}$- (or $C_{2}\times\mathcal{T}$-) symmetric (trivialized) fragile phases with corner modes was subsequently confirmed and expanded in Refs.~\onlinecite{KoreanFragile,WiederAxion,KoreanAxion,WladCorners,KoreanInversionFragile}.

\begin{figure}[h]
\centering
\includegraphics[width=0.90\textwidth]{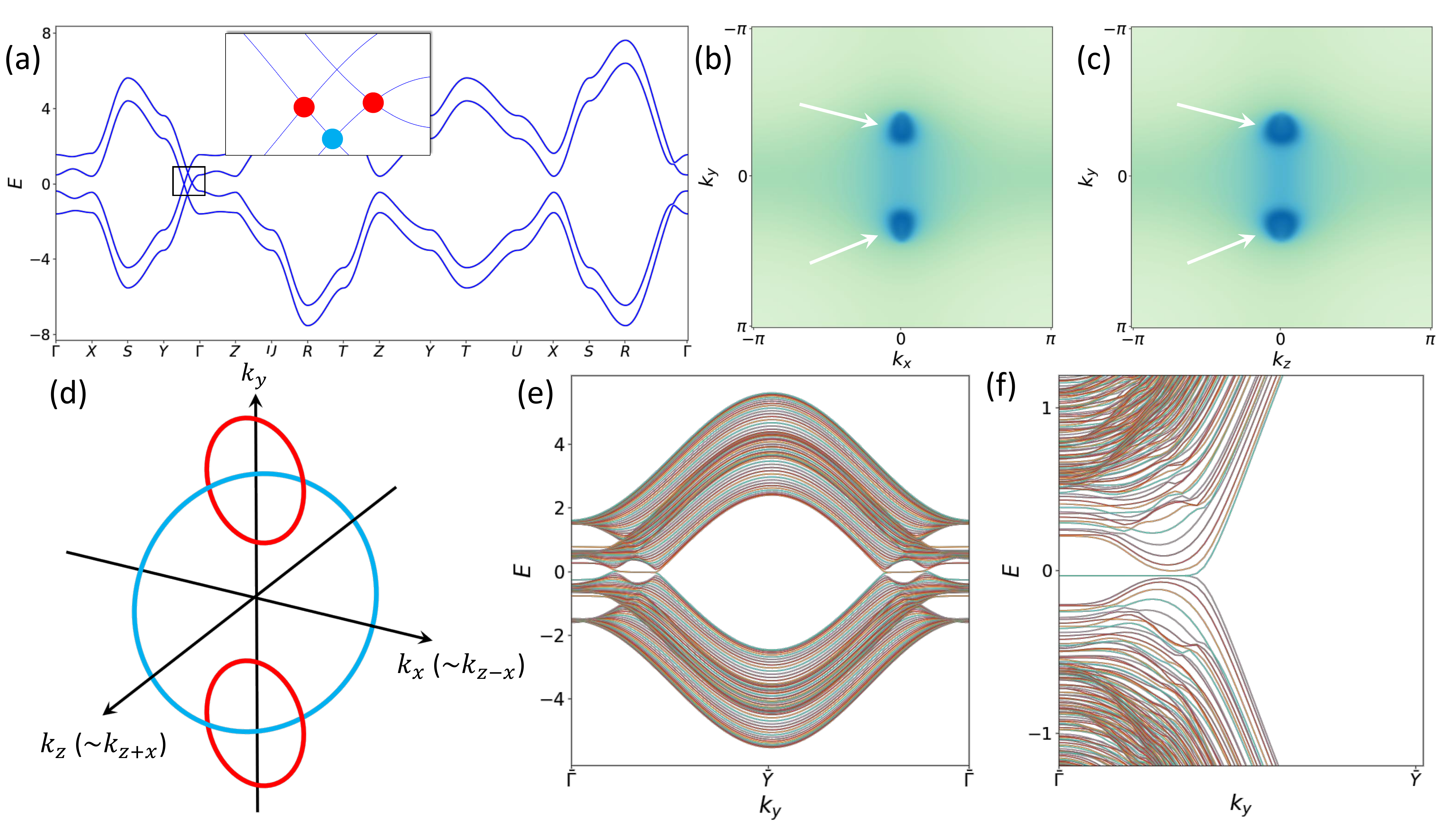}
\caption{(a) Bulk bands for $\tilde{\mathcal{H}}_{C}(\vec{k})$ in Eq.~(\ref{eq:coupledHam}), plotted with the parameters in Eqs.~(\ref{eq:uncoupledParams}) and~(\ref{eq:coupledParams}).  (b,c) $(001)$ and $(100)$ surface Green's function at $E_{F}=0$ for the same model, respectively.  Topological drumhead states appear on both surfaces in the interior projections of the MNLs (white arrows), indicating that the MNLs have become tilted and now have nonzero projections in both the $x$ and $z$ directions.  Crucially, the extraneous surface spectral weight spanning the projections of the MNLs from Fig.~\ref{fig:uncoupled}(d) has been lifted.  (d) Specifically, we observe that both the nodal lines at half filling (red), and the large nodal line directly below it in energy (blue) have become tilted by $\sim 45^\circ$ about the $k_{y}$ axis (axes in parenthesis are those after including Eq.~(\ref{eq:coupledHam})).  (e) The bands of a $z$-directed slab of $\tilde{\mathcal{H}}_{C}(\vec{k})$, plotted at $k_{x}=0$ as a function of $k_{y}$, confirm that the extra surface states have hybridized and split.  (f) The bands of a $y$-directed rod of $\tilde{\mathcal{H}}_{C}(\vec{k})$ (finite in the $x$ and $z$ directions).  Flat-band-like 1D states can be observed spanning the hinge projections of the MNLs; these are the spinless, $\mathcal{I}\times\tilde{\mathcal{T}}$-protected analogs of the hinge states recently predicted in spinful Dirac semimetals~\cite{HingeSM,TaylorToy}.}
\label{fig:coupled}
\end{figure}

\begin{figure}[h]
\centering
\includegraphics[width=\textwidth]{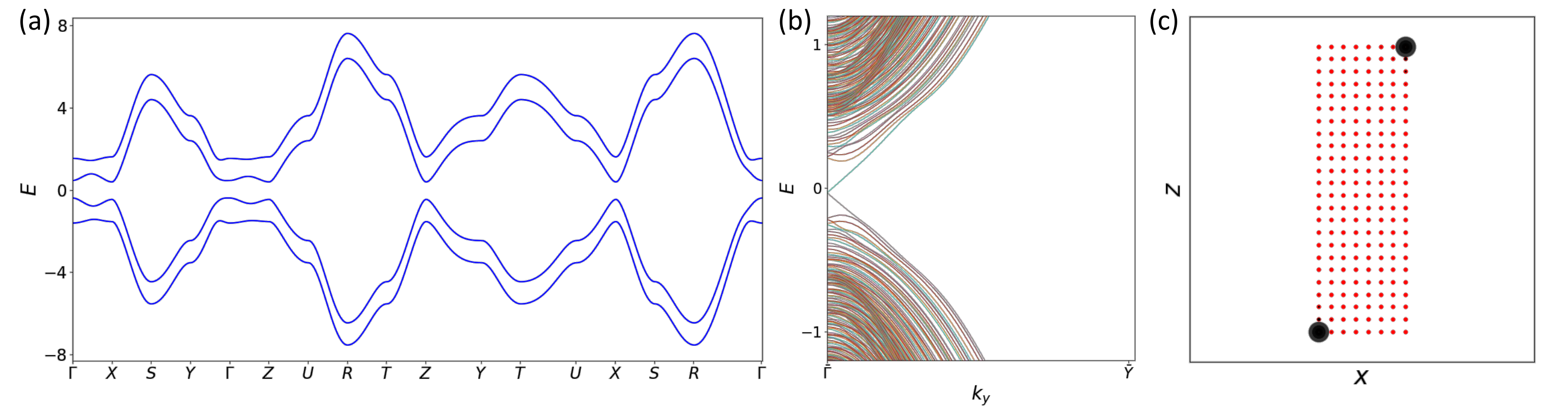}
\caption{(a) Bulk and (b) hinge bands of a $y$-directed rod of $\tilde{\mathcal{H}}_{C}(\vec{k})$ (Eq.~(\ref{eq:coupledHam})) with the Kane-Mele-like SOC term $V_{HOTI}(\vec{k})$ (Eq.~(\ref{eq:HOTI})), plotted with the parameters in Eqs.~(\ref{eq:uncoupledParams}) and~(\ref{eq:coupledParams}) and $v_{H}=1.2$.  The flat-band hinge states from Fig.~\ref{fig:coupled}(f) have evolved into a pair of 1D helical modes.  (c) The localization in the $xz$-plane of the hinge states at $k_{y}=0$; the two helical pairs of hinge modes are localized on $\mathcal{I}$-related hinges, confirming that $V_{H}(\vec{k})$ induces a phase transition from an MNL semimetal to an $\mathcal{I}$- and $\mathcal{T}$-symmetric HOTI~\cite{HigherOrderTIPiet,HigherOrderTIPiet2,HOTIBismuth,ChenTCI,AshvinIndicators,AshvinTCI,EslamInversion,WiederAxion}.  When $V_{Axion}(\vec{k})$ (Eq.~(\ref{eq:axion})) is used instead of $V_{HOTI}(\vec{k}))$, the bulk bands appear similar to those in (a), but the hinge spectrum instead exhibits oppositely propagating spin-degenerate pairs of chiral modes on $\mathcal{I}$-related hinges; this magnetic insulator is the spinless analog of the axion insulators analyzed in Refs.~\onlinecite{VDBHOTI,WiederAxion,KoreanAxion}.}
\label{fig:HOTI}
\end{figure}

To detect the hinge states implied by Eqs.~(\ref{eq:mainHamiltonianCircle}) and~(\ref{eq:nestedJackiw}) and the surrounding text, we must formulate a tight-binding model without the extraneous surface drumhead states shown with red arrows in Fig.~\ref{fig:uncoupled}(d,f), such that there is a projected bulk and surface gap in the spectrum of a $y$-directed rod~\cite{HingeSM}.  We first examine the extraneous surface spectral weight in Fig.~\ref{fig:uncoupled}(d) more closely by calculating the bands of a $z$-directed slab of $\tilde{\mathcal{H}}(\vec{k})$ with the parameters listed in Eq.~(\ref{eq:uncoupledParams}) (Fig.~\ref{fig:uncoupled}(f)).  We observe that the process of double band inversion has, in addition to nucleating the expected drumhead states in the interior projections of the MNLs (Fig.~\ref{fig:uncoupled}(d), white arrows), left behind a trivial pair of drumhead states~\cite{YoungkukLineNode,XiLineNode,SchnyderDrumhead} where the nodal line at half filling was pinched and split (Fig.~\ref{fig:uncoupled}(d,f), red arrows).  As there is no surface wallpaper group symmetry~\cite{DiracInsulator,WiederLayers} that protects the overlap of the extra surface states, we are free to add a bulk term that couples and gaps the trivial drumheads, analogous to the coupling that is naturally present in $\beta$-MoTe$_2$ (Fig.~\ref{fig:surf} of the main text).  We therefore introduce the bulk term:
\begin{equation}
\tilde{V}(\vec{k}) = m_{v1}\mu^{z} + m_{v2}\mu^{x},
\label{eq:couplingMass}
\end{equation}
realizing the coupled Hamiltonian:
\begin{equation}
\tilde{\mathcal{H}}_{C}(\vec{k}) = \tilde{\mathcal{H}}(\vec{k}) + \tilde{V}(\vec{k}).
\label{eq:coupledHam}
\end{equation}
Choosing the parameters:
\begin{equation}
m_{v1} = -0.4,\ m_{v2} = 0.2,
\label{eq:coupledParams}
\end{equation}
in addition to those listed in Eq.~(\ref{eq:uncoupledParams}), we again plot the bulk bands, $(001)$ surface Green's function, $(100)$ surface Green's function, and the $z$-directed slab bands at $k_{x}=0$ (Fig.~\ref{fig:coupled}(a,b,c,e)).  We observe that drumhead states now appear on both the $(001)$ and $(100)$ surfaces (Fig.~\ref{fig:coupled}(b,c), white arrows), indicating that the bulk MNLs have become tilted (Fig.~\ref{fig:coupled}(d)), and now carry nonzero interior projections in both the $x$ and $z$ directions.  Crucially, in the region between the surface projections of the bulk MNLs, the two extraneous drumhead states have become hybridized and split by the new mass terms in Eq.~(\ref{eq:couplingMass}).

Following the procedure employed in Ref.~\onlinecite{HingeSM}, we construct a $y$-directed rod of $\tilde{\mathcal{H}}_{C}(\vec{k})$, \emph{i.e.}, a tight-binding model that is finite in the $x$ and $z$ directions and infinite in the $y$ direction.  To understand the bulk and surface states that project to the hinges, one can take Fig.~\ref{fig:coupled}(b) and then project all of the surface spectral weight to the $k_{y}$ axis; the region between the two drumhead states (centered on the projection of $\Gamma$) remains free of surface (and bulk) states.  Plotting the hinge states of this rod (Fig.~\ref{fig:coupled}(f)), additional, 1D flat-band-like states are visible spanning the hinge projections of the MNLs.  Specifically, at each value of $k_{y}$ along the rod between the projections of the MNLs, there are four additional hinge states, which appear in spin-degenerate pairs localized on opposing hinges (Fig.~\ref{fig:HOTI}(c)).  These hinge states represent the $d-2$-dimensional generalization of the drumhead surface states of nodal-line semimetals~\cite{YoungkukLineNode,XiLineNode,SchnyderDrumhead}, and are the spinless analogs of the hinge states recently proposed in spinful Dirac semimetals~\cite{HingeSM,TaylorToy}.  It is clear that the $k_{y}$-indexed planes that exhibit hinge states in Fig.~\ref{fig:coupled}(f) lack the fourfold rotation and reflection symmetries of previously identified semimetals with hinge states~\cite{HingeSM,TaylorToy}.

In light of the relationship between MNLs and higher-order topology explored in the main text, we recognize the hinge states in Fig.~\ref{fig:coupled}(f) as the spinless precursors to the spinful helical hinge modes of 3D HOTIs.  They represent the higher-order generalization of the zigzag edge states of graphene~\cite{GrapheneReview,GrapheneEdge1,GrapheneEdge2,GrapheneEdgeMullen,GrapheneEdgeFan}, which analogously evolve into the helical edge modes of a 2D TI~\cite{KaneMeleZ2,CharlieTI} under the introduction of SOC.  As the MNLs are locally protected by $\mathcal{I}$, $\tilde{\mathcal{T}}$, and $SU(2)$ spin symmetry~\cite{YoungkukLineNode,FangWithWithout}, we can realize a bulk-insulating phase by relaxing one of these symmetries.  First, we reproduce the results of Ref.~\onlinecite{YoungkukMonopole} by introducing a term that breaks $\tilde{\mathcal{T}}$ symmetry while preserving $\mathcal{I}$ and $SU(2)$:
\begin{equation}
V_{Axion}(\vec{k}) = v_{A}\sin(k_{y})\tau^{y}\mu^{z}.
\label{eq:axion}
\end{equation}
We observe that $V_{Axion}(\vec{k})$ fully gaps the bulk and surface bands, realizing an insulating phase with spin-degenerate, chiral hinge modes (the bulk and hinge bands appear qualitatively similar to those shown in Fig.~\ref{fig:HOTI}, as bands of opposite chirality from opposing hinges become projected on top of each other).  Eq.~(\ref{eq:axion}) is therefore the higher-order analog of the magnetic Haldane term that gaps graphene into a spinless Chern insulator~\cite{HaldaneModel}.  We recognize the gapped 3D phase induced by $V_{Axion}(\vec{k})$ as exhibiting the same hinge spectrum (per spin) as the magnetic HOTI that results from gapping all of the surfaces of a 3D strong TI with magnetism that is spatially distributed in an $\mathcal{I}$-odd fashion~\cite{EzawaMagneticHOTI,AshvinTCI,VDBHOTI,WiederAxion}.  It is therefore also equivalent to two, spin-degenerate copies of a spinful axion insulator~\cite{WiederAxion,YoungkukMonopole} that are prevented from trivially hybridizing by $SU(2)$ symmetry.

Finally, we can also introduce a term of a similar form: 
\begin{equation}
V_{HOTI}(\vec{k}) = v_{H}\sin(k_{y})\tau^{y}\mu^{z}\sigma^{z},
\label{eq:HOTI}
\end{equation}
that breaks spinless time-reversal symmetry $\tilde{\mathcal{T}}$ and $SU(2)$ symmetry while preserving \emph{spinful} $\mathcal{T}$ symmetry, which is represented by the action:
\begin{equation}
\mathcal{T}\mathcal{H}(\vec{k})\mathcal{T}^{-1} = (i\sigma^{y}\tilde{\mathcal{T}})\mathcal{H}(\vec{k})(i\sigma^{y}\tilde{\mathcal{T}})^{-1} = \sigma^{y}\tau^{z}\mathcal{H}^{*}(-\vec{k})\sigma^{y}\tau^{z}.
\end{equation}
$V_{HOTI}$ term gaps the bulk MNLs (Fig.~\ref{fig:HOTI}(a)) and opens up the spin-degenerate flat-band hinge states (Fig.~\ref{fig:coupled}(e)) into the helical hinge modes of a HOTI.  Eq.~(\ref{eq:HOTI}) is therefore the higher-order analog of the Kane-Mele SOC term that gaps graphene into a 2D TI~\cite{KaneMeleZ2,CharlieTI}.  Thus, we have demonstrated that double band inversion in an $\mathcal{I}$- and $\mathcal{T}$-symmetric crystal with vanishing SOC can induce a pair of MNLs, which in turn can be gapped with $\mathcal{I}$-symmetric SOC to realize a $\mathbb{Z}_{4}$-nontrivial HOTI~\cite{AshvinIndicators,AshvinTCI,ChenTCI,EslamInversion}.

We note that, unlike MoTe$_2$ (Fig.~\ref{fig:surf}(d,e) of the main text and Fig.~\ref{fig:distort}(e,f) in Appendix~\ref{sec:DFTgappedGamma}), the HOTI induced by Eq.~(\ref{eq:HOTI}) \emph{does not} display large gapped surface Fermi arcs at low energies.  Instead, its bulk and surfaces are fully insulating (Fig.~\ref{fig:HOTI}(b)).  However, if parameters were adjusted to give greater dispersion to the gapped drumhead states near $\bar{\Gamma}$ in Fig.~\ref{fig:coupled}(e), then, in the presence of SOC (Eq.~(\ref{eq:HOTI})), the surface Green's functions in Fig.~\ref{fig:coupled}(b,c) would begin to exhibit pairs of arc-like states from gapped surface Dirac cones, like those in MoTe$_2$.  This reinforces the notion that, in $d$-dimensional insulators with higher-order topological boundary modes, the $d-1$-dimensional gapped surface states, while still topological (in the sense that they represent of anomalous ``halves'' of isolated $d-1$-dimensional topological (crystalline) insulators~\cite{HingeSM,WiederAxion,HOTIBernevig,DiracInsulator,WiederDefect,BarryConvo,BarryPrep}), can nevertheless be moved away from the Fermi energy (and possibly into the bulk manifolds) without breaking a symmetry or closing a bulk or surface gap.

\section{First-Principles Calculations Details}
\label{sec:DFT}

\subsection{Density Functional Theory Calculation Methods}
\label{sec:DFTmethod}

First-principles electronic structure calculations were performed with the projector augmented wave (PAW) method~\cite{paw1,paw2} as implemented in the VASP package~\cite{vasp1,vasp2}.  We adopted the Perdew-Burke-Ernzerhof (PBE) generalized gradient approximation (GGA) for the exchange-correlation functional~\cite{pbe}.  SOC was incorporated self-consistently.  The kinetic energy cutoff of the plane-wave basis was set to 400 eV. A 6$\times$12$\times$4 k-point mesh was employed for BZ sampling.  Internal atomic positions and cell parameters were obtained from experimental data for $\beta$-MoTe$_2$ in space group 11 $P2_{1}/m$ (ICSD~\cite{ICSD} \#14349)~\cite{Brown1966}.  The maximally localized Wannier functions (MLWF) were constructed from first-principles calculations~\cite{SMV2001} using the $d$ orbitals of Mo and the $p$ orbitals of Te.  The Wannier-based tight-binding Hamiltonian obtained from this construction was used to compute the surface spectrum and the nested Wilson loop~\cite{multipole,WladTheory,HOTIBernevig} matrix $W_{2}(k_{y})$ as described in Appendix~\ref{sec:DFTnested}.  

\subsection{First-Principles Calculation of Monopole Charge}
\label{sec:DFTmonopole}

When the effects of SOC are neglected, the electronic structure of $\beta$-MoTe$_2$ exhibits two nodal lines connecting the 28$^{\text{th}}$ and 29$^{\text{th}}$ spin-degenerate pair of bands, related by inversion symmetry, lying on either side of the $k_{y}=0$ plane, and intersecting $Y\Gamma$ (Fig.~\ref{fig:Fig2}(b) of the main text).  Rather than surround one of these nodal lines with a sphere, as is done in Refs.~\onlinecite{YoungkukMonopole,AdrianMonopole,SigristMonopole} for tight-binding models, we surround it with a geometrically simpler closed tetragonal prism.  Defining the Wilson matrix as the product of the adjacent overlap matrices -- $\braket{u_{{\bf k}_1}}{u_{{\bf k}_2}}$ --, where $\ket{u_{{\bf k}}}$ is the cell-periodic part of the Bloch eigenstate, we calculate the phases of the Wilson loop eigenvalues over the lower 28 spin-degenerate pairs of bands on the following paths along this prism, shown in Fig.~\ref{fig:Fig2}(c) of the main text.  We begin by calculating the loops on the bottom of the prism along the path $(x,0,x)\rightarrow(-x,0,x)\rightarrow(-x,0,-x)\rightarrow(x,0,-x)\rightarrow(x,0,x)$, taking $x$ to vary from $0$ to $0.45$ in units of the reciprocal lattice vectors.  We choose the bound $x=0.45$ such that the prism can contain as much as possible of the half BZ without touching the zone edge.  We then take loops of increasing height $y$ along the sides of the prism along the path $(0.45,y,0.45)\rightarrow(-0.45,y,0.45)\rightarrow(-0.45,0,-0.45)\rightarrow(0.45,0,-0.45)\rightarrow(0.45,0,0.45)$, taking $y$ to vary from $0$ to $0.5$.  We finally close the exterior of the prism by taking square loops on the top of decreasing width $2x$, where each loop is taken along the path $(x,0.5,x)\rightarrow(-x,0.5,x)\rightarrow(-x,0.5,-x)\rightarrow(x,0.5,-x)\rightarrow(x,0.5,x)$, taking $x$ to vary from $0.45$ to $0$.  We plot in the inset panel of Fig.~\ref{fig:Fig2}(b) of the main text the resulting Wilson spectrum as a function of $x$ for the bottom, then $y$ for the sides, and then finally $-x$ for the top, which we condense and label as the overall ``polar momentum'' $k_{\theta}$. The Wilson loop eigenvalues exhibit the characteristic winding of an MNL (Appendix~\ref{sec:TBmodel})~\cite{YoungkukMonopole,AdrianMonopole,SigristMonopole}.

\subsection{Calculating the Wilson Loop of the Wilson Loop and Quantization of the Nested Berry Phase}
\label{sec:DFTnested}

Here, we detail the calculations performed to obtain the determinant of the nested Wilson loop matrix $W_{2}(k_{y})$ (defined rigorously in Appendix~\ref{sec:WilsonDefs}) in Fig.~\ref{fig:wloop} of the main text.  We first, neglecting the effects of SOC, calculate the $k_{z}$-directed Wilson loop matrix $W_{1}(k_{x},k_{y})$ over the lower 28 spin-degenerate pairs of bands, which can be expressed as $\braket{u^0}{u^N}\braket{u^{N}}{u^{N-1}}\dots \braket{u^{1}}{u^{0}}$, where N is the discretized N-th $k$ point of the line: $(k_x,k_y,0)\rightarrow(k_x,k_y,2\pi)$. Diagonalizing the resulting Wilson loop matrix, we obtain the eigenvectors $\ket{\tilde{u}^n(k_x,k_y)}$ and eigenvalues $\xi_{1}^{n}(k_x,k_y)$ as functions of $(k_x,k_y)$, where $n$ is the Wilson band index. The eigenvalues $\xi_{1}^{n}(k_x,k_y)$ appear in the form $e^{i\theta_{1}^{n}(k_x,k_y)}$.  In Fig.~\ref{fig:wloop}(c,d) of the main text, we show the calculated values of $\theta_{1}^{n}(k_x,0)$ and $\theta_{1}^{n}(k_x,\pi)$, respectively, which we refer to as the Wilson bands.  In all of the $k_{y}$-indexed planes away from the MNLs, the Wilson bands are well separated by gaps in the Wilson spectrum at $\theta_{1} = \pm \pi/2$ ($0.25\times (2\pi)$, as expressed in the main text).  The system $H_{W_{1}}(k_x,k_{y})\equiv\sum_{|\theta_{1}^{n}|<(\pi/2)}\ket{\tilde{u}^n(k_x,k_{y})}\theta_{1}^{n}(k_x,k_{y}) \bra{\tilde{u}^n(k_x,k_{y})}$ resembles a 1D periodic Hamiltonian for fixed values of $k_y$, and its eigenstates can be used to calculate a second, nested Wilson loop matrix $W_{2}(k_{y})$ whose determinant is equal to $e^{i\gamma_{2}(k_{y})}$, where $\gamma_{2}(k_{y})$ is the nested Berry phase~\cite{multipole,WladTheory,HOTIBernevig,WiederAxion} of each plane indexed by $k_{y}$.  We compute the determinant of $W_{2}(k_{y})$ for all values of $k_{y}$, and observe that it is quantized at $\pm 1$ for all values of $k_{y}$ away from the MNLs, and jumps as the plane on which it is calculated passes fully over an MNL (Fig.~\ref{fig:wloop}(b) of the main text).

\subsubsection{Nested Berry Phase Quantization from $\mathcal{I}\times\mathcal{T}$ Symmetry}
\label{sec:WilsonDefs}

We note that the choice of Wilson energy interval employed for the nested Wilson loop calculations in this letter is different than that used in previous works~\cite{multipole,WladTheory,HingeSM}.  Specifically, in previous works, the Wilson spectrum was divided into halves between $\theta_{1}=0,\pi$ for nested Wilson loop calculations; here, we instead choose the particle-hole-symmetric interval $\theta_{1}\in [-\pi/2,\pi/2)$.  However, as long as the nested Wilson loop (and Berry phase) is calculated over the same Wilson interval for two different 2D insulators (or planes of the BZ), it can be used as a tool to compare their topology~\cite{multipole}.  

Furthermore, we discover in this letter that when the nested Berry phase $\gamma_{2}$ is calculated over this new choice of Wilson energies, it can be quantized without relying on mirror and fourfold rotation, as was previously required to quantize the nested Berry phase of the quadrupole insulators in Refs.~\onlinecite{multipole,HingeSM}.  In this section, we will show that, in particular, the combined antiunitary symmetry $\mathcal{I}\times\mathcal{T}$ is sufficient to quantize $\gamma_{2}$ when $W_{2}$ is calculated over a particle-hole symmetric set of Wilson bands.  We will find that a set of Wilson bands with quantized $\gamma_{2}$ does not need to lie specifically within the Wilson energy range employed in this letter ($\theta_{1}\in [-\pi/2,\pi/2)$), or even be contiguous in Wilson energy~\cite{WiederAxion}; the only restriction is that the set of Wilson bands within the nested Wilson projector returns to itself under the action of a Wilson particle-hole-symmetry that originates from bulk $\mathcal{I}\times\mathcal{T}$ symmetry.  The existence of this particle-hole symmetry in the Wilson spectrum was first derived in Ref.~\onlinecite{ArisInversion}; we reproduce its derivation here for convenience, and then use the result to demonstrate the $\mathbb{Z}_{2}$ quantization of $\gamma_{2}$.

To begin, we first consider a 3D orthorhombic crystal with a bulk Hamiltonian $\mathcal{H}(k_{x},k_{y},k_{z})$ that is invariant under $\mathcal{I}\times\mathcal{T}$, where $\mathcal{T}$ can represent either spinless ($\mathcal{T}^{2}=+1$) or spinful ($\mathcal{T}^{2}=-1$) time-reversal.  We then calculate the discretized $z$-directed Wilson loop as it is defined in Refs.~\onlinecite{ArisInversion,Cohomological,DiracInsulator}:
\begin{align}
\left[ W_{1(k_\perp,k_{z0})}  \right]_{nm} & \equiv \left[ \mathcal{P} e^{i \int_{k_{z0}}^{k_{z0}+2\pi} dk_z A_z(k_\perp,k_{z0})}\right]_{nm}  \nonumber\\
&\approx  \left[ \mathcal{P} e^{i \frac{2\pi}{N}\sum_{j=1}^N A_z(k_\perp,k_{z0}+ \frac{2\pi j}{N}) }\right]_{nm} \nonumber\\
\approx \langle u^n(k_\perp,k_{z0}+2\pi) |  \bigg[ \mathcal{P} \prod_{j=1}^N  P(k_\perp,k_{z0}+\frac{2\pi j}{N}) & \left( 1  - \frac{2\pi }{N} \partial_{k_z}|_{(k_\perp,k_{z0}+ \frac{2\pi j}{N}) }\right) P(k_\perp,k_{z0}+\frac{2\pi j}{N}) \bigg] |u^m(k_\perp,k_{z0})\rangle \nonumber\\
&\approx \langle u^n(k_\perp,k_{z0})|V(2\pi\hat{z})\Pi(k_\perp,k_{z0}) |u^m(k_\perp,k_{z0}) \rangle, 
\label{eq:wilsondisc}
\end{align}
where $k_\perp \equiv (k_x,k_y)$, $P(\mathbf{k})$ is the projector onto the occupied states (here the separated grouping of energy bands):
\begin{equation}
P(\mathbf{k}) = \sum_{n=1}^{n_{\rm occ}}|u^n(\mathbf{k})\rangle \langle u^n(\mathbf{k})|,
\label{eq:defproj}
\end{equation}
and where in the last line of Eq.~(\ref{eq:wilsondisc}), we have defined the ordered product of projectors,
\begin{equation} 
\Pi(k_\perp,k_{z0}) \equiv P(k_\perp,k_{z0}  + 2\pi)P(k_\perp,k_{z0} +\frac{2\pi (N-1)}{N})\cdots P(k_\perp,k_{z0}+\frac{2\pi }{N}).
\end{equation} 
The discretized loop defined by the product of projectors in Eq.~(\ref{eq:wilsondisc}) is closed by a sewing matrix:
\begin{equation}
\left[V(2\pi\hat{z})\right]_{nm} =  |u^n(k_\perp,k_{z0})\rangle\langle u^m(k_\perp,k_{z0}+2\pi)|,
\label{eq:defsew}
\end{equation}
that enforces the gauge and basepoint ($k_{z0}$) invariance of the eigenvalues of Eq.~(\ref{eq:wilsondisc})~\cite{ArisInversion,Cohomological,DiracInsulator}.  We then define a Hermitian ``Wilson Hamiltonian,''
\begin{equation}
\left[H_{W_{1(k_{z0})}}(k_{x},k_{y})\right]_{nm} = \left[-i\ln(W_{1(k_{x},k_{y},k_{z0})})\right]_{nm} \equiv H_{W_{1}}(k_{x},k_{y})= -i\ln(W_{1}(k_{x},k_{y})),
\label{eq:WilsonHam}
\end{equation}
where in the equivalence we define the less formal expressions for the $z$-directed Wilson Hamiltonian and loop with suppressed band indices used throughout this letter.  

From the analysis provided in Refs.~\onlinecite{Cohomological,DiracInsulator}, we recognize that the bulk symmetry $\mathcal{I}\times\mathcal{T}$ acts on $H_{W_{1}}(k_{x},k_{y})$ as an antiunitary particle-hole symmetry $\tilde{\Xi}$, but one that does not change the signs of $k_{x,y}$.  Specifically, because $\mathcal{I}\times\mathcal{T}$ does not change the direction of the product of projectors in Eq.~(\ref{eq:wilsondisc}), the action of $\mathcal{I}\times\mathcal{T}$ on $W_{1}(k_{x},k_{y})$ can simply be deduced from Eqs.~(\ref{eq:defproj}) and (\ref{eq:defsew}):
\begin{align}
(\mathcal{I}\times\mathcal{T})W_{1(k_{x},k_{y},k_{z0})}(\mathcal{I}\times\mathcal{T})^{-1} & = (\mathcal{I}\times\mathcal{T})V(2\pi\hat{z})\Pi(k_{x},k_{y},k_{z0} )(\mathcal{I}\times\mathcal{T})^{-1}  \nonumber \\
&  = UV^{*}(2\pi\hat{z})\Pi^{*}(k_{x},k_{y},k_{z0})U^{\dag} \nonumber \\
& = UW^{*}_{1(k_{x},k_{y},k_{z0})}U^{\dag},
\end{align}
where $[U,W^{*}_{1(k_{x},k_{y},k_{z0})}]=0$ for spinless electrons.  The Wilson Hamiltonian is therefore invariant under an antiunitary particle-hole symmetry, which we denote as $\tilde{\Xi}$, that leaves $k_{x,y}$ invariant:
\begin{equation}
\tilde{\Xi} H_{W_{1}}(k_{x},k_{y})\tilde{\Xi}^{-1} = -\tilde{U}(k_{x},k_{y})H^{*}_{W_{1}}(k_{x},k_{y})\tilde{U}^{\dag}(k_{x},k_{y}),
\label{eq:ITWilson1}
\end{equation}
for which one can choose $\tilde{U}(k_{x},k_{y})=\mathds{1}$ for spinless electrons without loss of generality.  Eq.~(\ref{eq:ITWilson1}) implies that for every Wilson eigenstate $|\tilde{u}^n(k_{x},k_{y},k_{z0})\rangle$ with eigenvalue $\theta^{n}_{1}(k_{x},k_{y})$, there is another eigenstate with eigenvalue $-\theta^{n}_{1}(k_{x},k_{y})$: 
\begin{equation}
\tilde{\Xi}|\tilde{u}^n(k_{x},k_{y},k_{z0})\rangle = \tilde{U}'(k_{x},k_{y},k_{z0}) (|\tilde{u}^n(k_{x},k_{y},k_{z0})\rangle)^{*},
\label{eq:XiOnN}
\end{equation}
where $\tilde{U}'(k_{x},k_{y},k_{z0})$ is the product of $\tilde{U}(k_{x},k_{y})$ in Eq.~(\ref{eq:ITWilson1}) and a $\mathbf{k}$-dependent unitary transformation that rotates the Wilson band index $n$, and is present for both spinful and spinless electrons.

As $H_{W_{1}}(k_{x},k_{y})$ is well-defined and generically gapped in the momentum-space regions for which $\mathcal{H}(k_{x},k_{y},k_{z})$ is gapped (for a sufficiently large number of occupied bands (Refs.~\onlinecite{ArisInversion,YoungkukMonopole} and Appendix~\ref{sec:TBmodel})), we can calculate the $x$-directed nested Wilson matrix $W_{2}(k_{y})$ by projecting onto half of the eigenstates of $H_{W_1}(k_{x},k_{y})$ and repeating the Wilson loop calculation in Eq.~(\ref{eq:wilsondisc}).  Formally, we define the $x$-directed nested Wilson loop:
\begin{align}
\left[ W_{2(k_{x0},k_{y},k_{z0})}  \right]_{nm} & \equiv \left[ \mathcal{P} e^{i \int_{k_{x0}}^{k_{x0}+2\pi} dk_x \tilde{A}_x(k_{x0},k_{y},k_{z0})}\right]_{nm}  \nonumber\\
&\approx  \left[ \mathcal{P} e^{i \frac{2\pi}{N}\sum_{j=1}^N \tilde{A}_x(k_{x0}+ \frac{2\pi j}{N},k_y,k_{z0}) }\right]_{nm} \nonumber\\
\approx \langle \tilde{u}^n(k_{x0}+2\pi,k_y,k_{z0}) |  \bigg[ \mathcal{P} \prod_{j=1}^N  \tilde{P}(k_{x0}+\frac{2\pi j}{N},k_y,k_{z0} ) &  \left( 1  - \frac{2\pi }{N} \partial_{k_x}|_{k_{x0}+ \frac{2\pi j}{N},k_y,k_{z0}) }\right)   \tilde{P}(k_{x0}+\frac{2\pi j}{N},k_y,k_{z0}) \bigg] |\tilde{u}^m(k_{x0},k_{y},k_{z0})\rangle \nonumber\\
&\approx \langle \tilde{u}^n(k_{x0},k_{y},k_{z0})|\tilde{V}(2\pi\hat{x})\tilde{\Pi}(k_{x0},k_{y},k_{z0} ) |\tilde{u}^m(k_{x0},k_{y},k_{z0}) \rangle, 
\label{eq:wilsondisc2}
\end{align}
where tildes indicate quantities obtained from the Wilson Hamiltonian (Eq.~(\ref{eq:WilsonHam})), $\tilde{P}(\mathbf{k})$ is the projector onto the occupied Wilson states (here the separated grouping of Wilson bands):
\begin{equation}
\tilde{P}(\mathbf{k}) = \sum_{n=1}^{\tilde{n}_{\rm occ}}|\tilde{u}^n(\mathbf{k})\rangle \langle \tilde{u}^n(\mathbf{k})|,
\label{eq:defproj2}
\end{equation}
and where in the last line of Eq.~(\ref{eq:wilsondisc2}), we have defined the ordered product of Wilson projectors,
\begin{equation} 
\tilde{\Pi}(k_{x0},k_{y},k_{z0}) \equiv \tilde{P}(k_{x0}  + 2\pi,k_y,k_{z0}) \tilde{P}(k_{x0} +\frac{2\pi (N-1)}{N},k_y,k_{z0})\cdots \tilde{P}(k_{x0}+\frac{2\pi }{N},k_y,k_{z0}).
\label{eq:defNestedProduct}
\end{equation} 
The discretized nested loop in Eq.~(\ref{eq:wilsondisc2}) is also closed by a sewing matrix:
\begin{equation}
\left[\tilde{V}(2\pi\hat{x})\right]_{nm} =  |\tilde{u}^n(k_{x0},k_{y},k_{z0})\rangle\langle \tilde{u}^m(k_{x0}+2\pi,k_y,k_{z0})|,
\label{eq:nestedSew}
\end{equation}
that enforces the basepoint ($k_{x0}$) independence of Eq.~(\ref{eq:wilsondisc2})~\cite{multipole,WladTheory} in the same manner that Eq.~(\ref{eq:defsew}) does for Eq.~(\ref{eq:wilsondisc}).  The eigenvalues of $W_{2(k_{x0},k_{y},k_{z0})}$ are gauge-independent~\cite{multipole,WladTheory} and take the form of phases $\exp(i\theta_{2}(k_{y}))$.  We can thus define a Hermitian ``nested Wilson Hamiltonian,''
\begin{equation}
\left[H_{W_{2(k_{x0},k_{z0})}}(k_{y})\right]_{nm} = \left[-i\ln(W_{2(k_{x0},k_{y},k_{z0})})\right]_{nm} \equiv  H_{W_{2}}(k_{y})   = -i\ln(W_{2}(k_{y})),
\label{eq:WilsonHam2}
\end{equation}
where in the equivalence, we define the less formal expressions for the $x$-directed nested Wilson Hamiltonian and loop with suppressed Wilson band indices used throughout this letter.  The eigenvalues of $H_{W_{2(k_{x0},k_{z0})}}(k_{y})$ take the form of real angles $\theta_{2}(k_{y})$, and we refer to the values of $\theta_{2}(k_{y})$ as ``nested Wilson energies.''  At each value of $k_{y}$, the sum of the nested Wilson energies modulo $2\pi$ is equal to the nested Berry phase:
\begin{equation}
\gamma_{2}(k_{y}) = \sum_{n=1}^{\tilde{n}_{\rm occ}}\theta^{n}_{2}(k_{y})\text{ mod }2\pi.
\label{eq:nestedSum}
\end{equation} 

We can now determine the action of $\tilde{\Xi}$ (and thus $\mathcal{I}\times\mathcal{T}$) on the nested Wilson loop.  Crucially, in order for $\tilde{\Xi}$ to be a symmetry of the nested Wilson loop, we must restrict to nested Wilson projectors $\tilde{P}(\mathbf{k})$ onto particle-hole conjugate pairs of Wilson bands, such that:
\begin{eqnarray}
\tilde{\Xi}\tilde{P}(k_{x},k_{y},k_{z0})\tilde{\Xi}^{-1} &=& \tilde{U}'(k_{x},k_{y},k_{z0})\tilde{P}^{*}(k_{x},k_{y},k_{z0})\tilde{U}'^{\dag}(k_{x},k_{y},k_{z0}) \nonumber \\
&=& \tilde{U}(k_{x},k_{y},k_{z0})\tilde{P}^{*}(k_{x},k_{y},k_{z0})\tilde{U}^{\dag}(k_{x},k_{y},k_{z0}),
\label{eq:ITNestedProj}
\end{eqnarray}
projects onto the same Wilson bands as $\tilde{P}(k_{x},k_{y},k_{z0})$, and where one can again take $\tilde{U}(k_{x},k_{y},k_{z0})=\mathds{1}$ for spinless electrons without loss of generality.  For the $x$-directed nested Wilson loop, $\tilde{\Xi}$ also does not change the direction of the product of projectors in Eq.~(\ref{eq:wilsondisc2}), and so the action of $\tilde{\Xi}$ (and thus $\mathcal{I}\times\mathcal{T}$) on $W_{2}(k_{y})$ also follows simply from Eqs.~(\ref{eq:defproj2}),~(\ref{eq:nestedSew}), and~(\ref{eq:ITNestedProj}):
\begin{align}
\tilde{\Xi}W_{2(k_{x0},k_{y},k_{z0})}\tilde{\Xi}^{-1} & = \tilde{\Xi}\tilde{V}(2\pi\hat{x})\tilde{\Pi}(k_{x0},k_{y},k_{z0} )\tilde{\Xi}^{-1}  \nonumber \\
&  =  \tilde{U}(k_{x0},k_{y},k_{z0})\tilde{V}^{*}(2\pi\hat{x})\tilde{\Pi}^{*}(k_{x0},k_{y},k_{z0})\tilde{U}^{\dag}(k_{x0},k_{y},k_{z0}) \nonumber \\
& = \tilde{U}(k_{x0},k_{y},k_{z0})W^{*}_{2(k_{x0},k_{y},k_{z0})}\tilde{U}^{\dag}(k_{x0},k_{y},k_{z0}).
\label{eq:W2ITEq}
\end{align}
The nested Wilson Hamiltonian is therefore invariant under an antiunitary particle-hole symmetry, which we denote as $\tilde{\tilde{\Xi}}$, that preserves the sign of $k_{y}$:
\begin{equation}
\tilde{\tilde{\Xi}} H_{W_{2}}(k_{y})\tilde{\tilde{\Xi}}^{-1} = -\tilde{\tilde{U}}(k_{y})H^{*}_{W_{2}}(k_{y})\tilde{\tilde{U}}^{\dag}(k_{y}),
\label{eq:ITWilson2}
\end{equation}
for which one can take $\tilde{\tilde{U}}(k_{y})=\mathds{1}$ for spinless electrons without loss of generality.  Eq.~(\ref{eq:ITWilson2}) implies that for every nested Wilson eigenstate $|\tilde{\tilde{u}}^n(k_{x0},k_{y},k_{z0})\rangle$ with eigenvalue $\theta^{n}_{2}(k_{y})$, there is another eigenstate with eigenvalue $-\theta^{n}_{2}(k_{y})$: 
\begin{equation}
\tilde{\tilde{\Xi}}|\tilde{\tilde{u}}^n(k_{x0},k_{y},k_{z0})\rangle = \tilde{\tilde{U}}'(k_{x0},k_{y},k_{z0}) (|\tilde{\tilde{u}}^n(k_{x0},k_{y},k_{z0})\rangle)^{*},
\label{eq:XiOnNW2}
\end{equation}
where $\tilde{\tilde{U}}'(k_{x0},k_{y},k_{z0})$ is the product of $\tilde{\tilde{U}}(k_{y})$ in Eq.~(\ref{eq:ITWilson2}) and a $\mathbf{k}$-dependent unitary transformation that rotates the nested Wilson band index $n$, and is present for both spinful and spinless electrons.  Taking the determinant of the right-hand side of Eq.~(\ref{eq:W2ITEq}) and exploiting that $W_{2}(k_{y})$ is invariant under $\tilde{\Xi}$:
\begin{equation}
\det(W_{2}(k_{y})) = \left(\det(W_{2}(k_{y}))\right)^{*},
\label{eq:QuantizedNestedBoth}
\end{equation}
for both spinful and spinless electrons.  Along with Eq.~(\ref{eq:nestedSum}), Eq.~(\ref{eq:QuantizedNestedBoth}) implies that at \emph{all} values of $k_{y}$ for which $W_{2}(k_{y})$ is well-defined (\emph{i.e.} there is a bulk and Wilson gap), $\det(W_{2}(k_{z}))$ is $\mathbb{Z}_{2}$ quantized:
\begin{equation}
\det(W_{2}(k_{y})) = \pm 1,
\end{equation}
and thus also implies that the nested Berry phase $\gamma_{2}(k_{y})$ is $\mathbb{Z}_{2}$ quantized:
\begin{equation}
\gamma_{2}(k_{y}) = 0,\pi. 
\label{eq:QuantIT}
\end{equation}

\subsection{Higher-Order Topology in Gapped $\gamma$-XTe$_2$}
\label{sec:DFTgappedGamma}

\begin{figure}[t]
\includegraphics[width=0.85\linewidth]{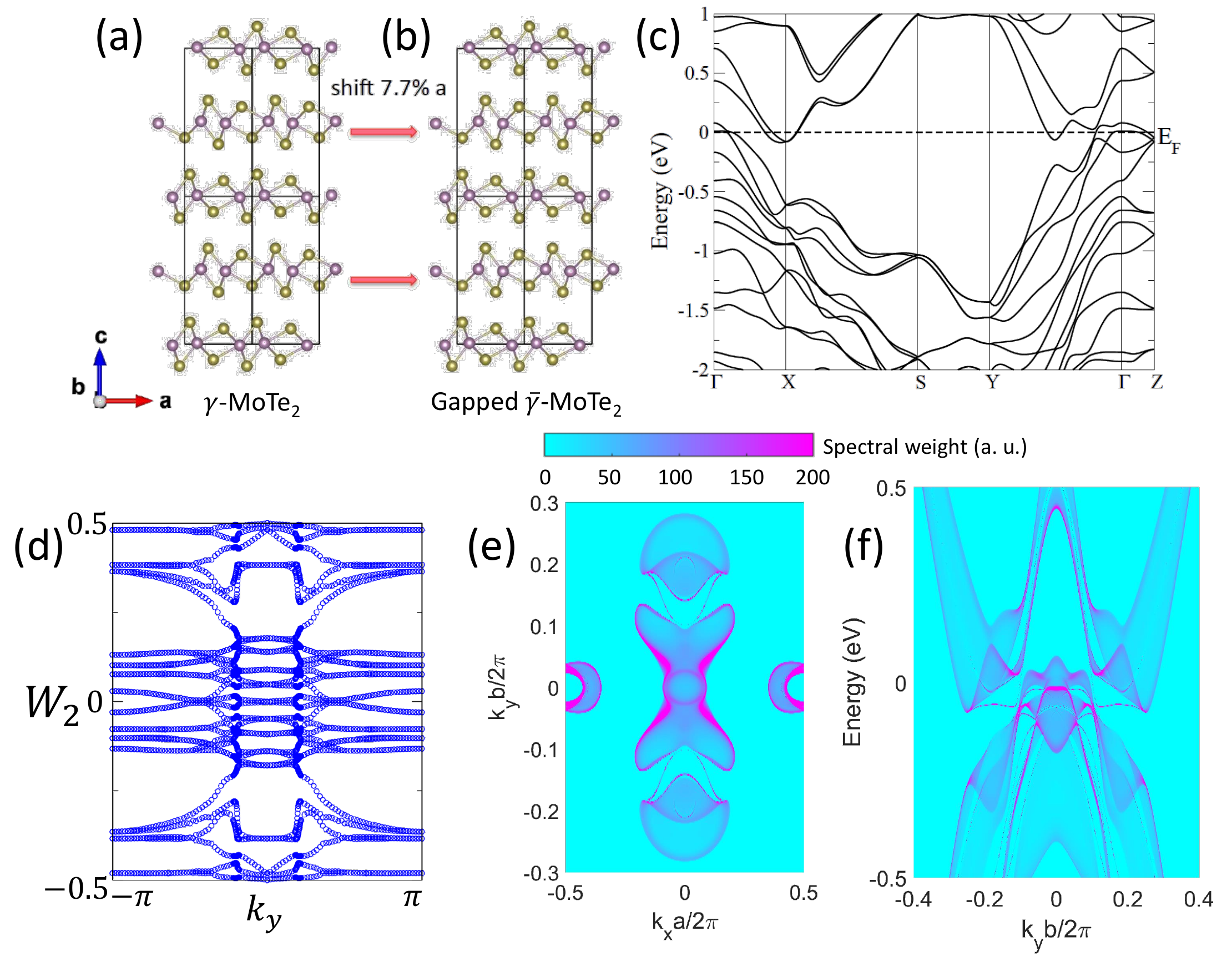}
\caption{(a) The crystal structure of $\gamma$-MoTe$_2$ (SG 31 in the nonstandard setting $Pnm2_{1}$, see text surrounding Eq.~(\ref{eq:nonstandard})) in the type-II Weyl semimetal phase studied in Ref.~\onlinecite{wang_mote2:_2016}.  (b) By sliding alternating layers of MoTe$_2$ along the $a$-axis by $\sim7.7 \%$ of the $a$-direction lattice spacing, we realize an artificial centrosymmetric structure in SG 62 $Pnma$, which we designate as $\bar{\gamma}$-MoTe$_2$.  The unstable $\bar{\gamma}$ structure coincides with the ``T$_{0}$'' structure of MoTe$_2$ introduced in Ref.~\onlinecite{DavidMoTe2} after the submission of this letter.  (c) The bulk band structure of $\bar{\gamma}$-MoTe$_2$ is fully gapped, and exhibits the same parity eigenvalues as $\beta$-MoTe$_2$ (Table~\ref{tb:z4} and Fig.~\ref{fig:Fig2}(d) of the main text).  We then calculate the $z$-directed Wilson loop $W_{1}(k_{x},k_{y})$ of $\bar{\gamma}$-MoTe$_2$, which, like the Wilson spectrum of $\beta$-MoTe$_2$ (Fig.~\ref{fig:wloop}(c,d) of the main text) is gapped at $\theta_{1}=\pm \pi/2$, indicating that $\bar{\gamma}$-MoTe$_2$ is a HOTI, and not a mirror TCI.  (d) $W_{2}(k_{y})$ of $\bar{\gamma}$-MoTe$_2$, calculated using the Wilson bands near $\theta_{1}=0$.  As shown in Appendix~\ref{sec:WilsonDefs}, $W_{2}(k_{y})$ is Wilson particle-hole symmetric at each value at $k_{y}$ due to the combined symmetry $\mathcal{I}\times\mathcal{T}$.  By generalizing the arguments for $\mathcal{I}$- and rotation-symmetry-protected HOTIs introduced in Refs.~\onlinecite{WiederAxion,HOTIBernevig,HOTIBismuth}, we conclude that the helical winding in (d) is representative of a strong topological phase enforced by $\mathcal{I}$ and $\mathcal{T}$ symmetries~\cite{WiederDefect,BarryPrep}, in agreement with the bulk nontrivial $\mathbb{Z}_{4}$ parity index.  As discussed in this section (Appendix~\ref{sec:DFTgappedGamma}), this also implies that $W_{2}(k_{z})$, which we found to be numerically difficult to calculate, must nevertheless be well-defined, and must also exhibit robust helical winding enforced by $\mathcal{I}$ and $\mathcal{T}$.  Furthermore, the implied helical winding of $W_{2}(k_{z})$ can alternatively be considered enforced by twofold screw and $\mathcal{T}$ symmetries~\cite{ChenTCI,WiederAxion}, which are also symmetries of $\gamma$-MoTe$_2$, unlike $\mathcal{I}$ symmetry.  Thus, $W_{2}(k_{z})$ should continue to exhibit strong helical winding under a slight distortion from $\bar{\gamma}$-MoTe$_2$ back to a gapped $\gamma$-phase (the reverse of (a) and (b)).  Therefore, gapped $\gamma$-MoTe$_2$ is a non-symmetry-indicated HOTI.  (e) Spectral weight at the Fermi energy of states on the $(001)$ surface of $\bar{\gamma}$-MoTe$_2$, calculated using the same methodology employed for Fig.~\ref{fig:surf}(d,e) of the main text (Appendix~\ref{sec:DFTmethod}), and plotted as a function of the in-plane momenta $k_{x,y}$, and (f) along $k_{x}=0$ as a function of energy.  The surface states of $\bar{\gamma}$-MoTe$_2$ are gapped (f), and are nearly identical to the large, gapped, nontrivial, arc-like HOTI surface states of $\beta$-MoTe$_2$ (Fig.~\ref{fig:surf}(d,e) of the main text).  Because the surface states in (d,e) will remain gapped under infinitesimal distortion from $\bar{\gamma}$-MoTe$_2$ to gapped $\gamma$-MoTe$_2$, this provides further evidence that $\gamma$-MoTe$_2$ is a noncentrosymmetric, non-symmetry-indicated HOTI when gapped.}
\label{fig:distort}
\end{figure}

In this section, we explicitly show that $\gamma$-MoTe$_2$, when gapped, is a noncentrosymmetric, non-symmetry-indicated HOTI.  We begin with $\gamma$-MoTe$_{2}$, calculated with the structural parameters used in Ref.~\onlinecite{wang_mote2:_2016} (Fig.~\ref{fig:distort}(a)).  In the $\gamma$ structure, MoTe$_2$ crystals are left invariant under the action of space group (SG) 31 in the nonstandard setting of $Pnm2_{1}$, which is generated by:
\begin{equation}
n_{x} = \bigg\{M_{x}\bigg|0\frac{1}{2}\frac{1}{2}\bigg\},\ m_{y} =\bigg\{M_{y}\bigg|000\bigg\},\ s_{2_{1}z} = \bigg\{C_{2z}\bigg|0\frac{1}{2}\frac{1}{2}\bigg\},
\label{eq:nonstandard}
\end{equation}
as obtained from the~\href{http://www.cryst.ehu.es/cryst/get_gen.html}{GET GEN} tool on the Bilbao Crystallographic Server~\cite{BCS1}.  Specifically, SG 31 is typically associated (and is listed for simplicity in the main text) with the standard symbol~\cite{BigBook} $Pmn2_{1}$.  However, because the standard symbol implies the (symmorphic) mirror reflection $m_{x}=\{M_{x}|000\}$, whereas the actual mirror symmetry in $\gamma$-MoTe$_2$ is $m_{y}=\{M_{y}|000\}$ (Eq.~(\ref{eq:nonstandard})), then in this section we will use the more precise nonstandard symbol $Pnm2_{1}$ to characterize the structure of $\gamma$-MoTe$_2$ in SG 31.  When the bulk band structure of $\gamma$-MoTe$_2$ is calculated with the structural parameters used in Ref.~\onlinecite{wang_mote2:_2016}, it exhibits tilted (type-II) Weyl points~\cite{soluyanov_type-ii_2015,wang_mote2:_2016}, and is thus semimetallic, and not insulating.  

To gap the Weyl points, we slide alternating layers of $\gamma$-MoTe$_2$ along the $a$-axis by $\sim7.7\%$ of the $a$ lattice spacing (Fig.~\ref{fig:distort}(b)).  After this distortion, the crystal develops an artificial, unstable centrosymmetry:
\begin{equation}
\mathcal{I} = \bigg\{\mathcal{I}\bigg|\frac{1}{2}\frac{1}{2}0\bigg\},
\label{eq:fakeInversion}
\end{equation}
and is thus characterized by SG 62 $Pnma$, a supergroup~\cite{BigBook} of SG 31 $Pnm2_{1}$ that is generated by adding Eq.~(\ref{eq:fakeInversion}) to the SG 31 generators in~\cite{BigBook,BCS1} Eq.~(\ref{eq:nonstandard}).  The new, artificial structure in SG 62 $Pnma$, which we denote as $\bar{\gamma}$-MoTe$_2$ (Fig.~\ref{fig:distort}(b)), coincides with the unstable ``T$_{0}$'' structure, which was introduced in Ref.~\onlinecite{DavidMoTe2} after the submission of this letter.  Plotting the bulk band structure of $\bar{\gamma}$-MoTe$_2$ (Fig.~\ref{fig:distort}(c)), we observe that all of the Weyl points along high symmetry lines have been removed.  Furthermore, because nonmagnetic crystals in SG 62 $Pnma$ host the combined symmetry $\mathcal{I}\times\mathcal{T}$, all of their bands are at least twofold degenerate, and hence cannot meet in conventional Weyl points along the lower-symmetry BZ planes and interior~\cite{Vafek14} not pictured in Fig.~\ref{fig:distort}(c).  Therefore, $\bar{\gamma}$-MoTe$_2$ is a bulk band insulator (though still metallic) at the Fermi energy.

Next, we diagnose the bulk topology of $\bar{\gamma}$-MoTe$_2$.  The bulk bands exhibit all of the same parity eigenvalues as $\beta$-MoTe$_2$ (Table~\ref{tb:z4} and Fig.~\ref{fig:Fig2}(d) of the main text).  Thus, using the $\mathbb{Z}_{4}$ parity index developed in Refs.~\onlinecite{ChenTCI,AshvinIndicators,AshvinTCI,EslamInversion} and in the main text, $\bar{\gamma}$-MoTe$_2$ exhibits the strong indices of an $\mathcal{I}$- and $\mathcal{T}$-symmetric HOTI.  We use Wilson loops to further determine that the mirror Chern numbers~\cite{TeoFuKaneTCI,HsiehTCI} of the $y$-directed mirrors are trivial in both the $k_{y}=0,\pi$ planes.  Specifically, we calculate the $z$-directed Wilson loop $W_{1}(k_{x},k_{y})$, using the 56 highest valence bands in energy, and find that it is gapped in the vicinity of $\theta_{1}=\pm \pi/2$ at all values of $k_{x,y}$ (producing a Wilson spectrum similar to that of $\beta$-MoTe$_2$ shown in Fig.~\ref{fig:wloop}(c,d) of the main text), indicating that the $y$-directed mirror Chern numbers are trivial.  Therefore, $\bar{\gamma}$-MoTe$_2$ is a HOTI, and not a TCI, as is also allowed by its bulk symmetry eigenvalues~\cite{ChenTCI,AshvinIndicators,AshvinTCI}.  We further confirm the bulk topology by calculating the $x$-directed nested Wilson loop $W_{2}(k_{y})$ of the separated grouping of Wilson bands near $\theta_{1}=0$ (Fig.~\ref{fig:distort}(d)), employing the same numerical methods previously used for $\beta$-MoTe$_2$ (Appendix~\ref{sec:DFTnested}).  As shown Appendix~\ref{sec:WilsonDefs}, $W_{2}(k_{y})$ is Wilson particle-hole symmetric at each value of $k_{y}$ due to the combined symmetry $\mathcal{I}\times\mathcal{T}$.  We observe that $W_{2}(k_{y})$ exhibits helical winding (Fig.~\ref{fig:distort}(d)).  Generalizing the arguments presented in Refs.~\onlinecite{HOTIBernevig,WiederAxion} for rotation- and $\mathcal{I}$-protected HOTIs, band crossings in $W_{1}(k_{x},k_{y})$ can only manifest in Wilson particle-hole and $\mathcal{T}$-symmetric pairs, and thus the winding of $W_{2}(k_{y})$ cannot be removed by a gap closure in the Wilson spectrum that is not accompanied by a gapless point in the energy spectrum.  Therefore, the helical winding of $W_{2}(k_{y})$ shown in Fig.~\ref{fig:distort}(d) is indicative of a strong (higher-order) topological phase.  A more rigorous $\mathcal{T}$-symmetric generalization of nested Berry phase will appear in Ref.~\onlinecite{WiederDefect}.

Furthermore, in any $\mathcal{I}$- and $\mathcal{T}$-symmetric HOTI with gapped surface states and a large number of occupied bands~\cite{WiederAxion}, any nested Wilson loop, performed as prescribed in Appendix~\ref{sec:DFTnested}, must exhibit helical winding.  Therefore, $W_{2}(k_{z})$, which we found numerically difficult to explicitly calculate due the more complicated forms of the $x$- and $y$-directed Wilson bands in $\bar{\gamma}$-MoTe$_2$, must nevertheless be well-defined, and display helical winding if correctly computed.  Though it has not yet been explicitly demonstrated, we here outline how this winding can alternatively be interpreted as protected by the combination of twofold screw ($s_{2_{1}z}$) and $\mathcal{T}$ symmetries, leaving the formal details for future works.  In Refs.~\onlinecite{WiederAxion,KoreanAxion}, it was shown that the combined magnetic symmetry of twofold rotation $C_{2z}$ and $\mathcal{T}$ protects strong, odd-integer chiral winding of a nested Wilson loop $W_{2}(k_{z})$ directed along the rotation axis ($z$-direction), and indicates that the bulk is a non-symmetry-indicated axion insulator.  Because the simplest HOTI can be formed from superposing two time-reversed axion insulators~\cite{HOTIBernevig,WiederAxion}, and because the irreducible (co)representations of $C_{2z}$ and $s_{2_{1}z}$ are closely related~\cite{BigBook,QuantumChemistry,Bandrep1}, then it is straightforward to argue that the presence of both $s_{2_{1}z}$ and $\mathcal{T}$ symmetries in a 3D insulator can enforce helical winding of $W_{2}(k_{z})$ that is indicative of strong, higher-order (crystalline) topology.  Because $\gamma$-MoTe$_{2}$ is noncentrosymmetric, and thus cannot exhibit strong (nested) Wilson loop winding enforced by $\mathcal{I}$ symmetry, we will find this reasoning crucial for arguing that it nevertheless exhibits the bulk band-insulating topology of an $s_{2_{1}z}$- and $\mathcal{T}$-protected HOTI.

To further support our diagnosis of $\bar{\gamma}$-MoTe$_2$ as a HOTI, we calculate the $(001)$ surface states through surface Green's functions (Fig.~\ref{fig:distort}(d,e)).  Of the bulk crystal symmetries of SG 62 $Pnma$ (Eqs.~(\ref{eq:nonstandard}) and~(\ref{eq:fakeInversion})), only $M_{y}$ (as well as the rectangular lattice translations $T_{x,y}$) are preserved on the $(001)$ surface.  Therefore, the $(001)$ surface of $\bar{\gamma}$-MoTe$_2$ respects wallpaper group~\cite{DiracInsulator,WiederLayers,SteveMagnet} $pm$, which can only support twofold linear degeneracies at TRIM points and along the $m_{y}$-invariant BZ lines $k_{y}=0,\pi$.  Because the bulk $\mathbb{Z}_{4}$ invariant is even, then the strong Fu-Kane invariant is necessarily trivial~\cite{HOTIBernevig,HOTIBismuth,ChenTCI,AshvinIndicators,AshvinTCI}, indicating that $\bar{\gamma}$-MoTe$_2$ is not a 3D TI, and does not host topological (unpaired) surface states at the $(001)$ surface TRIM points.  Next, because the surface states of $\bar{\gamma}$-MoTe$_2$, like those of $\beta$-MoTe$_2$, originate from bulk double band inversion at $\Gamma$ (Fig.~\ref{fig:distort}(c)), then mirror TCI cones~\cite{TeoFuKaneTCI,HsiehTCI} can only appear along $k_{y}=0$, which coincides with the $(001)$-surface projection of $\Gamma$.  In Fig.~\ref{fig:distort}(e), we plot the $(001)$ surface states of $\bar{\gamma}$-MoTe$_2$ at the Fermi energy, and observe the clear absence of mirror TCI cones along $k_{y}=0$, in agreement with our earlier observation that the $z$-directed Wilson loop $W_{1}(k_{x},0)$ of $\bar{\gamma}$-MoTe$_2$ is gapped.  We additionally calculate the surface states along $k_{x}=0$ as a function of energy (Fig.~\ref{fig:distort}(f)), further demonstrating that the $(001)$ surface states of $\bar{\gamma}$-MoTe$_2$ are gapped, like those of $\beta$-MoTe$_2$ (Fig.~\ref{fig:surf}(e) of the main text).  This confirms, in agreement with the (nested) Wilson loop calculations performed earlier in this section (Fig.~\ref{fig:distort}(d)), that $\bar{\gamma}$-MoTe$_2$ is not a TCI, but is instead a HOTI with large, gapped surface states.

With these symmetry arguments established, and with the results of the (nested) Wilson loop and surface-state calculations in Fig.~\ref{fig:distort}(c-e), we can finally exploit the previous diagnosis of $\bar{\gamma}$-MoTe$_2$ as a HOTI to diagnose the topology of gapped $\gamma$-MoTe$_2$.  We model the transition away from the artifical $\bar{\gamma}$ structure by infinitesimally sliding back alternating layers of MoTe$_2$ along the $a$-axis (the reverse of the process depicted in Fig.~\ref{fig:distort}(a,b)).  This sliding lowers the crystal symmetry from SG 62 $Pnma$ back to SG 31 $Pnm2_{1}$, the SG of $\gamma$-MoTe$_2$.  Because the sliding is infinitessimal, it does not close a bulk or surface gap, and therefore the bulk remains band insulating, and the $(001)$ surface continues to exhibit large, nontrivial, gapped surface states.  As shown in Ref.~\onlinecite{ChenTCI} using position-space symmetry arguments, nonmagnetic crystals in SG 31 $Pnm2_{1}$ can only host three kinds of topological surface or hinge states: the surface cones of a strong 3D TI, the surface cones of a mirror TCI, and the hinge states of an $s_{2_{1}z}$- and $\mathcal{T}$-protected ``rotation-anomaly''~\cite{ChenRotation,ChenTCI,WiederAxion} TCI, which we consider to be a form of HOTI.  Because $\bar{\gamma}$-MoTe$_2$ was previously determined to be neither a strong 3D TI nor a mirror TCI, and because we previously determined that its nested Wilson loop $W_{2}(k_{z})$ must helically wind, then we conclude that gapped $\gamma$-MoTe$_2$ is a non-symmetry-indicated HOTI whose higher-order topology is protected by twofold screw and $\mathcal{T}$ symmetries.  Furthermore, because the electronic and crystal structure of $\gamma$-WTe$_2$ is nearly identical to that of~\cite{UnstableBetaWTe2} $\gamma$-MoTe$_2$, then we conclude that $\gamma$-WTe$_2$, when gapped, is also a non-symmetry-indicated HOTI.

\end{appendix}
\bibliography{hoti_mote2Arx5}
\end{document}